\documentclass[iop,apj]{emulateapj}
\usepackage{amsmath,amssymb,color}

\newcommand{\3}{I\hspace{-.1em}I\hspace{-.1em}I}
\newcommand{\2}{I\hspace{-.1em}I}

\newcommand{\Rc}{R_{\rm{c}}}
\newcommand{\Mc}{M_{\rm{c}}}
\newcommand{\Ec}{E_{\rm{c}}}
\newcommand{\vc}{v_{\rm{c}}}
\newcommand{\Rph}{R_{\rm{ph}}}

\newcommand{\Msun}{M_\odot}

\begin{document}
\title{Jet-powered supernovae of $\sim 10^5\Msun$ Population \3 stars are observable by \textit{Euclid}, \textit{WFIRST}, \textit{WISH}, and \textit{JWST}}

\author{Tatsuya Matsumoto\altaffilmark{1}, Daisuke Nakauchi\altaffilmark{2}, Kunihito Ioka\altaffilmark{3,4,5} and Takashi Nakamura\altaffilmark{1}}
\altaffiltext{1}{Department of Physics, Kyoto University, Kyoto 606-8502, Japan}
\altaffiltext{2}{Astronomical Institute, Tohoku University, Aoba, Sendai 980-8578, Japan}
\altaffiltext{3}{Theory Center, Institute of Particle and Nuclear Studies, KEK, Tsukuba 305-0801, Japan}
\altaffiltext{4}{Department of Particle and Nuclear Physics, SOKENDAI (The Graduate University for Advanced Studies), Tsukuba 305-0801, Japan}
\altaffiltext{5}{Yukawa Institute for Theoretical Physics, Kyoto University, Kyoto, 606-8502, Japan}
\begin{abstract}
Supermassive black holes observed at high redshift $z\gtrsim6$ could grow from direct collapse black holes (DCBHs) with mass $\sim10^5\Msun$, which result from the collapse of supermassive stars (SMSs).
If a relativistic jet is launched from a DCBH, it can break out of the collapsing SMS and produce a gamma-ray burst (GRB).
Although most of the GRB jets are off-axis from our line of sight,
we show that the energy injected from the jet into a cocoon is huge $\sim10^{55-56}\,{\rm{erg}}$, so that the cocoon fireball is observed as ultra-luminous supernovae of $\sim10^{45-46}\rm{\,erg\,s^{-1}}$ for $\sim5000 [(1+z)/16] \rm{\,days}$.
They are detectable by the future telescopes with near infrared bands, such as, \textit{Euclid}, \textit{WFIRST}, \textit{WISH}, and \textit{JWST} up to $z\sim20$ and $\lesssim10$ events per year,
providing a direct evidence of the DCBH scenario.
\end{abstract}


\section{Introduction}\label{Introduction}
Since the last decade, supermassive black holes (SMBHs) with $\sim10^{9}\,\Msun$ have been discovered in the high-redshift quasars~(QSOs) at $z\gtrsim6$ \citep{2006NewAR..50..665F,2011Natur.474..616M,2015Natur.518..512W}.
The origin of these SMBHs is one of the biggest riddles in the Universe.
We do not know how the seed BHs acquire mass of $\sim10^{9}\,\Msun$ within a short time of $\lesssim 1\ {\rm Gyr}$~(the age of the Universe at $z\gtrsim6$).

Plausible candidates for the seeds are stellar mass BHs of $\sim10^{1-3}\,\Msun$, which are the end products of first stars or Population \3~(Pop \3) stars.
Theoretical studies have shown that Pop \3 stars are typically very massive \citep[$\sim10^{2-3}\,\Msun$;][]{1999ApJ...527L...5B,2002ApJ...564...23B,2002Sci...295...93A,2003ApJ...592..645Y,2006ApJ...652....6Y,2008Sci...321..669Y,2014ApJ...792...32S,2014ApJ...781...60H,2015arXiv151001407H}.
Stellar mass seeds are difficult to grow to
be supermassive within $\lesssim 1\ {\rm Gyr}$ as long as the mass accretion continues with the Eddington rate.
Such continuous and efficient accretion may be prevented by the radiation feedback effects \citep{2009ApJ...698..766M,2009ApJ...701L.133A,2011ApJ...739....2P,2012ApJ...747....9P,2013ApJ...767..163P,2014ApJ...797..139A}.
Some authors suggest that supercritical accretion may help the stellar mass BHs to grow rapidly \citep{2005ApJ...633..624V,2014ApJ...784L..38M,2015arXiv151102116I,2014Sci...345.1330A}.

A more attractive seed may be provided by the supermassive Pop \3 stars~(SMSs) of $\sim10^5\,\Msun$.
SMSs are formed in primordial gas clouds which are under intense far ultra-violet (FUV) radiations \citep{2001ApJ...546..635O,2003ApJ...596...34B,2008MNRAS.391.1961D,2012MNRAS.425.2854A,2014MNRAS.445..686J,2015arXiv150404042A,2015MNRAS.446..160A,2015MNRAS.454.2441S} or in the high density and high temperature regions which are formed through the cold inflow onto a protogalaxy and/or galaxy mergers \citep{2012MNRAS.422.2539I,2015MNRAS.453.1692I}.
When SMSs collapse via the general relativistic (GR) instability \citep{1964ApJ...140..417C} or the exhaustion of the nuclear fuel, they leave massive BHs of $\sim10^5\ \Msun$ as remnants. We call these massive BHs as direct collapse BHs~(DCBHs).
With the Eddington accretion rate, such massive seeds can grow up to SMBHs of $\sim 10^9\ \Msun$ for $\sim 0.5\ {\rm Gyr}$.
This is shorter than the age of the Universe at $z = 7$, where the most distant QSO is found~\citep{2011Natur.474..616M}.

The detection of violent explosions, like supernovae (SNe) and gamma-ray bursts~(GRBs), produced by SMSs may be useful to understand their contribution to the SMBH formation \citep{2013ApJ...775..107J,2013ApJ...777...99W,2013ApJ...778...17W,2014ApJ...790..162C}.
In the previous paper, we studied whether SMSs can produce GRBs or not, and discussed their detectability \citep{2015ApJ...810...64M}.
GRBs are produced by the relativistic jets which are launched from BH-accretion disk systems.
In the collapsar scenario, the BH and disk system are formed in the center of a massive star, when it collapses  \citep{1993ApJ...405..273W,1999ApJ...524..262M} \citep[see also, ][for Pop \3 collapsars]{2000ApJ...536....1L,2004ApJ...604..508G,2010ApJ...715..967M,2011ApJ...726..107S,2011A&A...533A..32D,2012ApJ...754...85N,2012ApJ...759..128N,2014ApJ...787...91M}.
After the relativistic jet breaks out of the progenitor envelope, it can contribute to the $\gamma$- and X-ray prompt emission.
If the direction of the jet axis coincides with our line of sight, we can observe it as a GRB.

The stellar evolution theory suggests that SMSs have very large radii of $10^{14-15}$ cm, which are comparable to or even larger than red supergiants \citep{2011AN....332..408F, 2012ApJ...756...93H,2013ApJ...778..178H,2015MNRAS.452..755S}.
Although one may think that the bloated envelope prevents the relativistic jet from breaking out of it successfully, we have found that the breakout is possible because of the steeply-declining density profile.
We have also found that the GRBs from SMSs show ultra-long durations of $\delta{t_\gamma}\sim10^{4-6}\,\rm{s}$, which are about $10^{3-5}$ times longer than ordinary GRBs and even longer than the observed ultra-long GRBs~(ULGRBs\footnote{So far, ULGRBs are discovered in the low-$z$ Universe of $z \lesssim 1$ \citep{2013ApJ...766...30G,2014ApJ...781...13L}.
They have typical durations of $\delta{t}_{\gamma} \sim 10^4\,\rm{s}$ and the isotropically radiated energy of $\sim 10^{53}\ {\rm erg}$.
}).
The isotropically radiated energy of the GRBs amounts to as much as $E_{\gamma,\rm{iso}}\sim10^{56}\rm{\,erg}$, which is also by an order of magnitude larger than that of the most energetic GRB.
They could be detectable by the Burst Alert Telescope (BAT) onboard the \textit{Swift} satellite \citep{2005SSRv..120..143B} up to $z=20$.
Thus, very energetic ULGRBs
are a unique  feature of the GRBs from SMSs. 

However, the prompt emission has some disadvantages.
First, the detection rate of such GRBs was estimated to be low $\lesssim 2\ (\Psi_{\rm GRB}/10^{-8} {\rm yr}^{-1} {\rm Mpc}^{-3}) (\theta/5^{\circ})^2\ {\rm yr}^{-1}$ on the whole sky, where $\Psi_{\rm GRB}$ is the intrinsic event rate of the SMS GRBs, and $\theta$ is the opening angle of the jet.
Second, we cannot measure the distance or redshift only through the high energy emission, because it is difficult to identify the host galaxy.
In order to overcome these disadvantages, we consider the counterparts of SMS GRBs which radiate isotropically in the optical or near infrared~(NIR) bands.

In the previous study, we find that energy of $10^{55-56}\,{\rm erg}$ is injected into the hot plasma cocoon which surrounds a jet before its breakout \citep{2015ApJ...810...64M}.
When the jet propagates in the progenitor envelope, it forms two shock waves at the jet head.
One is a forward shock which sweeps the envelope materials, and the other is a reverse shock which decelerates the jet materials.
The materials in the jet head can expand sideways and form a hot plasma cocoon around the jet \citep{2003MNRAS.345..575M}.
The cocoon can also expand in the envelope and finally breaks out of it along with the jet head. 
We call the cocoon component emerging out of the progenitor star as a cocoon fireball.

Some authors suggested that the cocoon fireball evolves like an SN ejecta, if the cocoon loads the stellar material efficiently before breakout \citep{2013ApJ...770....8K,2013ApJ...778...67N}.
\cite{2013ApJ...778...67N} showed that the super-luminous SN~(SLSN) associated with the ULGRB 111209A can be reproduced by the emission from the cocoon fireball, if the progenitor is a blue supergiant of $\sim 10^{13}\ {\rm cm}$.
For SMSs, it takes much longer time for the jet head to break out of the envelope, so that we can expect much larger energy stored in the cocoon and much brighter emission from the cocoon fireball.

In this paper, we study the cocoon emission associated with the ULGRBs from SMSs.
We find that they can be observed as ultra-luminous SNe up to $z \sim 20$ by the future NIR surveys.
We also show that the distance or redshift of the event can be identified by the Gunn-Peterson trough \citep{1965ApJ...142.1633G}.
Furthermore, we find how to estimate the mass of the progenitor, its radius, and the explosion energy from the observables, such as
the bolometric luminosity, duration, and photospheric velocity.
This enables us to confirm that the progenitor is a SMS.
Finally, we discuss that from the detection rate by the future NIR surveys, we can constrain the conditions and environments for the SMS formation.

We organize this paper as follows.
In Section \ref{Cocoon emission from supermassive stars}, we show the light curves of the cocoon emission and discuss their detectability with future telescopes.
In Section \ref{Event rate}, we consider the observational strategy for the cocoon emission and their event rate.
In Appendix, we collect the formulae for calculating the cocoon parameters and the light curves.
Throughout this paper, we consider the $\Lambda${CDM} cosmology and adopt the cosmological parameters as :
$H_0=67.8\,\rm{km\ s^{-1}\,Mpc^{-1}}$, $\Omega_{\rm{m}}=0.308$ and
$\Omega_{\Lambda}=0.692$~\citep{2015arXiv150201589P}.
 
\section{Ultra-luminous SNe from Supermassive collapsars}\label{Cocoon emission from supermassive stars}

\begin{table}[!t]
\begin{center}
\caption{Parameters of cocoon fireballs}
\label{cocoon parameter table}
\begin{tabular}{ccc}
\tableline\tableline
Progenitor Model & 1E5 & Accreting \\
\tableline
$\Ec$ [erg] & $1.0\times10^{56}$ & $4.2\times10^{55}$\\
$\Mc$ [$\Msun$] & $1.1\times10^{3}$ & $3.5\times10^{2}$\\
$\Rc(0)$ [cm] & $1.2\times10^{14}$ & $2.8\times10^{15}$\\
\tableline
\end{tabular}
\end{center}
{\bf{Notes.}}
The energy $\Ec$, mass $\Mc$, and initial radius $\Rc(0)$ of cocoon fireballs are shown. The values of the energy $E_{\rm{c}}$ and mass $M_{\rm{c}}$ are evaluated at the jet breakout.
The cocoon initial radius $R_{\rm{c}}(0)$ means the cocoon radius at the start of homologous expansion (see Appendix \ref{Analytical estimate for cocoon parameters} and \ref{Light curve model} for the definition of these quantities). The 1E5 model is a SMS which evolves from a metal-free ZAMS star. The Accreting model is a massive protostar growing under rapid-mass accretion of $1\,\Msun\,\rm{yr^{-1}}$ (see the text).
\end{table}

In this section, first we briefly explain the SMS formation and our progenitor models.
Next, we show the light curves of the cocoon emission comparing with other SN events.
Then, we show the light curves for the future telescopes and discuss the detectability.

While hydrogen molecules are the main coolant in a primordial gas cloud, they can be destroyed via photodissociation or collisional dissociation when the cloud is irradiated with the intense FUV field or is in a highly shock-compressed region.
Such a gas cloud can contract by its self-gravity only via hydrogen atomic cooling, so that the temperature of the cloud is kept high~($\sim10^4\,\rm{K}$).
A protostar formed in the center of the contracting cloud accretes the surrounding gas with a high accretion rate of $\dot{M}\sim0.1-1\,\Msun\,\rm{yr^{-1}}$.
If the accretion can continue over the stellar lifetime $\sim$ Myr, the protostar can become a SMS of $\sim10^5\,\Msun$ \citep[see also,][]{2008ApJ...682..745W,2009MNRAS.396..343R,2009MNRAS.393..858R,2010MNRAS.402.1249S,2011MNRAS.418..838W,2013MNRAS.433.1607L,2013A&A...558A..59S,2014MNRAS.439.1160R,2014MNRAS.445.1549I,2015MNRAS.446.2380B}.

We consider two models of SMSs in this work \citep[for more details of our models, see][]{2015ApJ...810...64M}.
First, if the accretion is halted by e.g., radiation feedback when the star obtains $\sim 10^5\,\Msun$, the SMS reaches the zero age main sequence~(ZAMS).
\cite{2011AN....332..408F} calculated the evolution of a SMS of $10^5\,\Msun$ from the ZAMS stage until it exhausts its nuclear fuels and begins to collapse.
We use the density profile at the precollapse phase as a progenitor model, and call it as the ``1E5 model''.
Second, if the accretion continues without any interruption, they grow up to the critical mass where the GR instability sets in \citep{1964ApJ...140..417C}.
Then the accreting SMS will become unstable and finally collapse.
\cite{2013ApJ...778..178H} calculated the evolution of an accreting SMS with the constant and high accretion rate of $\dot{M}=1\ \Msun\,\rm{yr^{-1}}$.
They stopped the calculation when the star obtains $10^5\,\Msun$, since they suffered from some numerical difficulties.
While the GR instability has not set in yet, we adopt the density profile of this phase as a progenitor model, and call it as the ``Accreting model''.
As long as the accretion continues, the envelope profile will not change so much and it should have little effect on our results.

It should be noted that our progenitor models are calculated without taking rotation into account.
In reality, GRB progenitors are thought to be rotating for the jet formation.
When the progenitors are rotating very rapidly, the stellar structure may also become chemically homogeneous \citep{2005A&A...443..643Y,2006ApJ...637..914W}.
This may change the radii of the progenitors.
However, as long as the progenitor envelopes are radiation-pressure-dominated and have steeply-decreasing density profiles, the jet heads do not decelerate and can penetrate the stellar surface (see Appendix \ref{Analytical estimate for cocoon parameters}, for the jet propagation in a radiation-pressure-dominated envelope).

First, we calculate the dynamics of a relativistic jet in a progenitor envelope, and figure out the total energy $\Ec$ and mass $\Mc$ loaded on a cocoon.
We show the details of our model in Appendix A, along with the order of magnitude estimates.
In Table \ref{cocoon parameter table}, we show the results of the energy $\Ec$, mass $\Mc$ at the jet breakout, and initial radius $\Rc(0)$ of the cocoon fireballs when homologous expansion starts (see also Appendix \ref{Light curve model}).
The quantities $\Ec$ and $\Mc$ can be roughly reproduced by the analytical estimates~(Eqs. \ref{cocoon energy2} and \ref{cocoon mass2}). 
We find that the energy of the cocoon fireball is very large $\Ec \sim 10^{55-56}\ {\rm erg}$.
This is because the progenitors have a very large radius.
It takes much time for the jet head to break out of the progenitor envelope, so that a large amount energy is stored in the cocoon.
This can be seen from Eqs. \eqref{jet breakout time} and \eqref{cocoon energy2}.
We can see that the cocoon fireballs are non-relativistic, $\Ec\ll\Mc{c^2}$, and initially optically thick, $\tau_0 \sim\kappa\Mc/\Rc(0)^2\gg1$, where $c$ is the speed of light and $\kappa=0.35\rm{\,cm^2\,g^{-1}}$ is the Thomson scattering opacity for the primordial composition.
Thus, the cocoon fireball may evolve like a shock heated ejecta of Type \2P SNe \citep{2013ApJ...770....8K,2013ApJ...778...67N}.

\begin{figure}[!t]
\centering
\includegraphics[scale=0.3,angle=270]{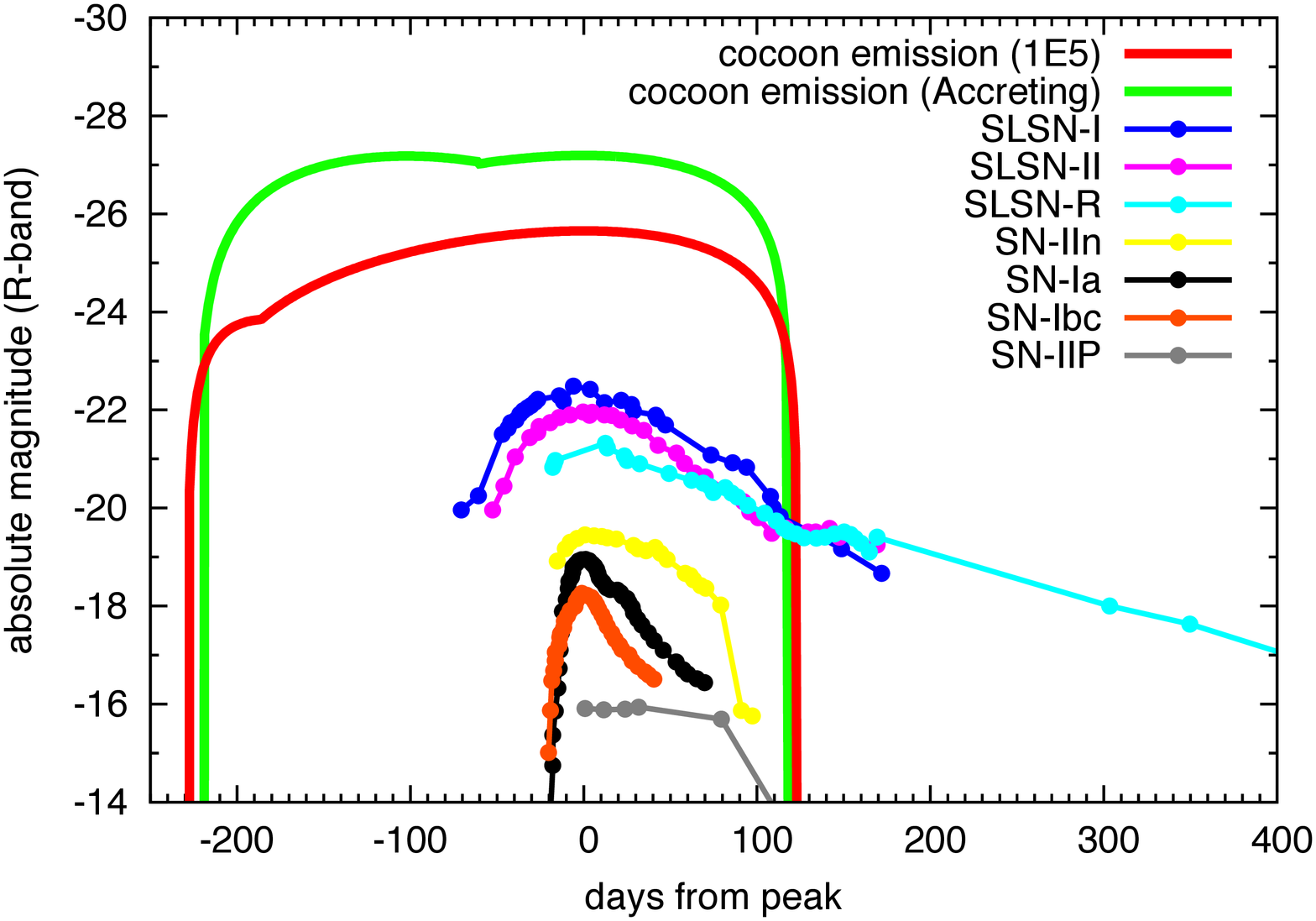}
\caption{Light curves of cocoon emission and other observed SNe. The horizontal axis shows days from their peaks in the light curves (For the cocoon emission, we represent the time in the progenitor frame). The vertical axis represents the absolute magnitude in the R-band. The red and green curves are the light curves of the cocoon emission for the 1E5 and Accreting models, respectively. The blue, magenta, light-blue, yellow, black, orange, and grey curves show the light curves of observed transient events, Type-I super-luminous SN (SLSN-I) PTF09cnd, Type-\2 SLSN SN2006gy , Type-R SLSN SN2007bi, Type \2n SN 2005cl, Type Ia SN, Type Ibc SN, and Type \2P SN 1999em, respectively. These light curves are taken from Fig. 1 in \cite{2012Sci...337..927G}.}
\label{cocoon_compare image}
\end{figure}

Next, using the parameter values in Table \ref{cocoon parameter table}, we calculate the light curves of the cocoon emission.
We show the details of our prescription in Appendix \ref{Light curve model}.
In Fig. \ref{cocoon_compare image}, we compare the light curves of the cocoon emission obtained from the 1E5 and Accreting model with those of the observed SNe.
The horizontal axis shows the time from the peak of the light curve (for the cocoon emission, we represent the time in the progenitor frame).
The vertical axis gives the absolute magnitude in the R-band.
The red and green solid curves correspond to the 1E5 model and the Accreting model, respectively.
We see that the cocoon emission are about $10-100$ times brighter than even SLSNe.
Its bolometric luminosity amounts to $10^{45-46}\ \rm{\,erg\,s^{-1}}$~(Eq. \ref{cocoon luminosity1}), so that we call them as ultra-luminous SNe. 
Such ultra-luminous SNe may be useful to study the early Universe.

From the green solid line in Fig. \ref{cocoon_compare image}~(the Accreting SMS model), we find that the light curve can be divided into two parts, and that each part has its own peak.
They come from the different thermal states in the cocoon fireball (Appendix \ref{Light curve model}).
Around the first peak the ejecta has very high temperature, so that the atoms are completely ionized.
In this phase, the effective temperature decreases with time according to Eq. \eqref{effective temperature}, so that the spectral peak moves to the redder bands with time. \if{~(see e.g., Fig. \ref{lc_1e5_wish2 image})}\fi
When the effective temperature drops below the critical value~($T_{\rm{ion}} \sim 6000$ K), the atoms recombine in the ejecta.
In the recombined ejecta, photons can escape almost freely, so that the recombination front becomes the photosphere.
Since the recombination front moves inwards to the expanding ejecta, the photospheric radius looks almost unchanged in the lab-frame observer (Eq. \ref{photospheric radius in recombination}).
Then, the spectral energy distribution hardly changes with time, and the light curve shows a plateau.
This phase corresponds to the plateau phase in Type \2P SNe.
The light curves of the cocoon emission drop abruptly at $\sim\,120$ days from the peaks.
This is because the photosphere reaches the center of the cocoon fireball and all photons diffuse out from the ejecta.

\if{ The light curves of the cocoon emission drop abruptly at $\sim\,120$ days from the peaks, while the light curves of SNe have a tail getting dark slowly.
These tails form when SN ejecta is heated by the radioactive decay of synthesized $^{56}\rm{Ni}$ \& $^{56}\rm{Co}$ \citep{1982ApJ...253..785A}.
On the other hand, cocoon fireballs are composed of heated stellar materials and do not contain heat sources such as $^{56}$Ni \citep{2013ApJ...770....8K}.
Therefore, the cocoon emission shows an abrupt darkening as its photosphere reaches the center.
Even if the light curves of the cocoon emission have a similar tail as SNe, the tails will not so much	 bright that we can observe them with telescopes discussed bellow.}\fi

\begin{table*}[!t]
\begin{center}
\caption{Properties of future NIR telescopes}
\label{telescope property table}
\begin{tabular}{cccc}
\tableline\tableline
Telescope & \textit{Euclid} & \textit{WFIRST} & \textit{WISH} \\
\tableline
Band & Y,J,H & Y,J,H,F184 & 1-4.5$\,\mu\rm{m}$ \\
Depth(SN survey) [mag] &26 & $29.3$-$29.4$(J,H\footnote{\textit{WFIRST} is planned to have three types of SN surveys \citep{2013arXiv1305.5422S}. In this paper, we only consider the SN deep survey, because it has the best sensitivity. The SN deep survey uses only J- and H-bands.}) & $-$ \\
Area(SN survey) [deg$^2$] & $40$ & $5.04$ & $-$\\
Duration(SN survey) [yr] & $3$ & $0.5$ & $-$\\ 
Cadence [day] & $4$-$6$ & $5$ & $-$\\
Depth(galaxy survey) [mag] & 24 & $26.2$-$26.9$ & $28$ \\
Area(galaxy survey) [deg$^2$] & $1.5$-$2.0\times10^{4}$ & $2.0\times10^{3}$ & $100$\\
Duration(galaxy survey) [yr] & $3$ & $1.3$ & $5$\\
\tableline
\end{tabular}
\end{center}
{\bf{Notes.}}
Area in line 4 and 8 means the size of the observed region by each telescopes.
In SN survey, telescopes observe the same regions many times and detect transient events.
In galaxy survey, telescopes survey large areas just once and observe galaxies.
\end{table*}

\begin{table}[!t]
\begin{center}
\caption{Center wavelength and maximum redshift of each band}
\label{band table}
\begin{tabular}{ccccccc}
\tableline
Band name & Y & J & H & F184 & K & L\\
wavelength [$\mu{\rm m}$] &1.020&1.215&1.654&1.842&2.179&3.547\\
$z_{\rm{max}}$ &7.4&9.0&12.6&14.2 &18.6&28.2\\ 
\tableline
\end{tabular}
\end{center}
{\bf{Notes.}}
In line 2, we show the center wavelength of each band.
At the maximum redshift of $z_{\rm{max}}$ in line 3, the center wavelength is equal to the redshifted Lyman-$\alpha$ wavelength, $\lambda_{\rm{Band}}=\lambda_{\rm{Ly}\alpha}(1+z_{\rm{max}})$.
\end{table}

Finally, we discuss the detectability of the cocoon emission with the future wide-field NIR survey telescopes, such as \textit{Euclid} \citep{2011arXiv1110.3193L}, \textit{WFIRST}  \citep{2013arXiv1305.5422S}, and \textit{WISH}\footnote{http://www.wishmission.org/en/index.html} (see also section \ref{Event rate}).
We show the property of each telescope in Table \ref{telescope property table}.
We set the redshifts of the events as $z=10$, $15$, and $20$, where SMSs are suggested to be formed \citep{2012MNRAS.425.2854A,2014MNRAS.442.2036D,2014MNRAS.440.1263Y}.
At these redshifts, intergalactic hydrogen will be still neutral \citep{2015arXiv150201589P}, so that the emission with wavelengths shorter than $\lambda_{\rm{obs}} = 0.122(1+z)\,\rm{\mu{m}}$ is strongly absorbed.
Therefore, we can use only redder bands than the redshifted Lyman-$\alpha$ wavelength.
In Table \ref{band table}, we show, for each band, the maximum redshift up to which photons at the band center are free from Lyman-$\alpha$ absorption.

\begin{figure}[!t]
\centering
\includegraphics[scale=0.3,angle=270]{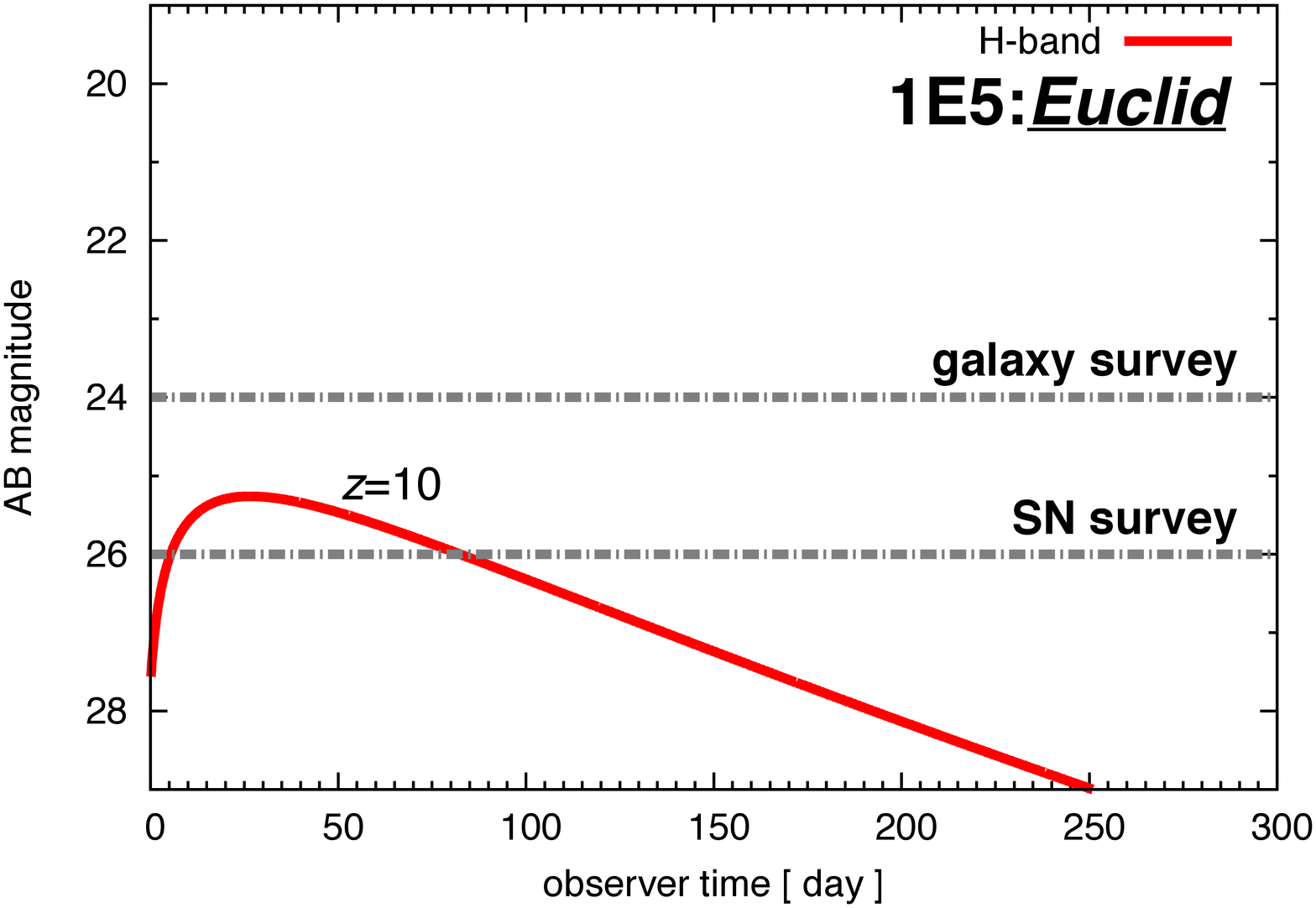}
\caption{Light curves of cocoon emission for the 1E5 SMSs observed with \textit{Euclid}. The horizontal axis shows the observer time since the cocoon fireball starts homologous expansion. The vertical axis shows the observed AB magnitude. The red curve shows the light curve in H-band. We show the sensitivities of \textit{Euclid}'s SN survey and galaxy survey with horizontal grey dash-dotted lines. \textit{Euclid} can detect only the first peaks of the cocoon emission in the SN survey.}
\label{lc_1e5_euclid image}
\end{figure}

\begin{figure}[!t]
\centering
\includegraphics[scale=0.3,angle=270]{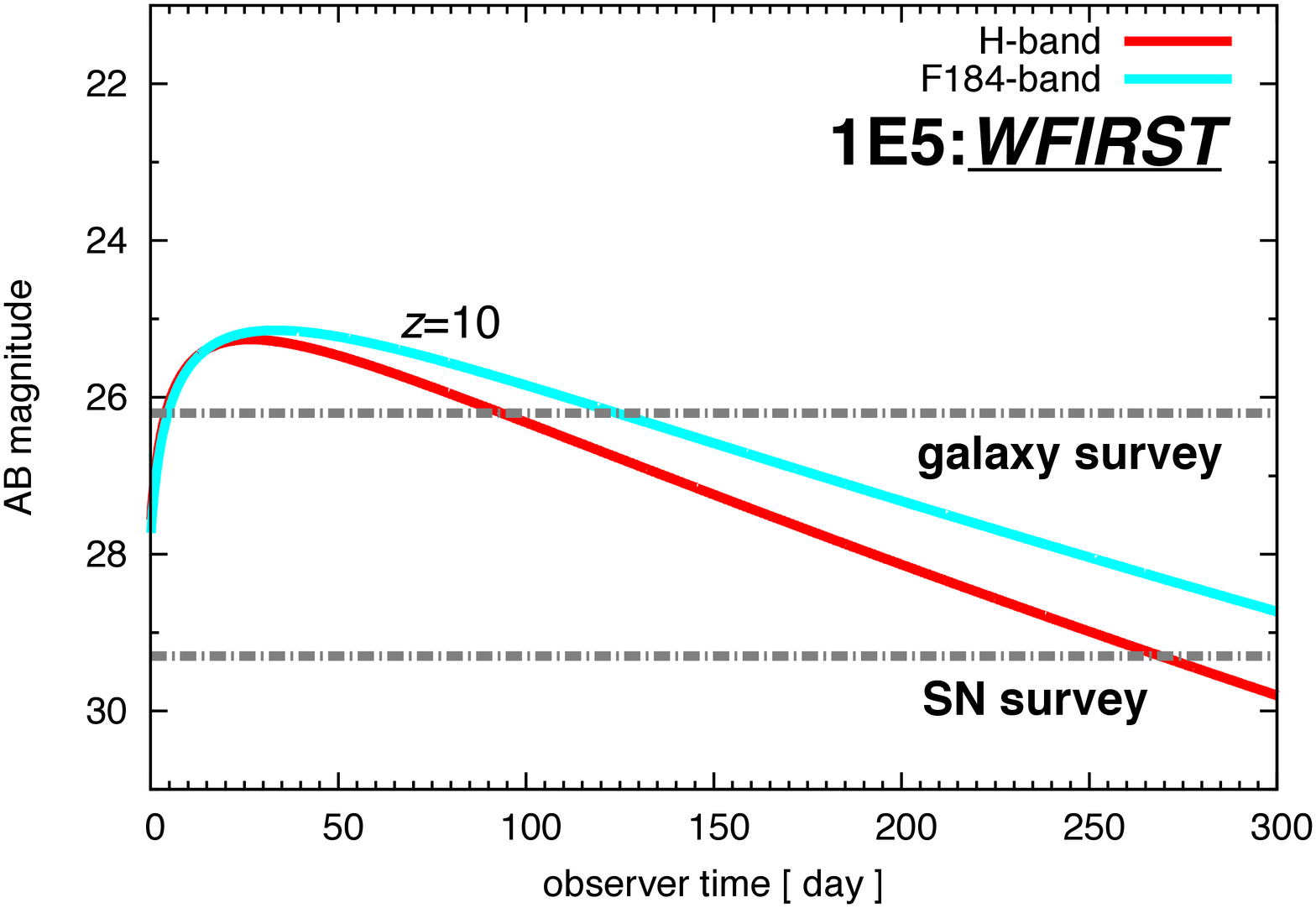}
\caption{Same as Fig. \ref{lc_1e5_euclid image}, but for the 1E5 SMSs observed with \textit{WFIRST}.
We add the light curves in F184-band with a light-blue curve.
\textit{WFIRST} can observe only the first peaks of the cocoon emission at $z=10$ for more than $90\rm{\,{days}}$.
It should be noted that in SN survey, \textit{WFIRST} uses only J- and H-bands.}
\label{lc_1e5_wfirst image}
\end{figure}

\begin{figure}[!t]
\centering
\includegraphics[scale=0.3,angle=270]{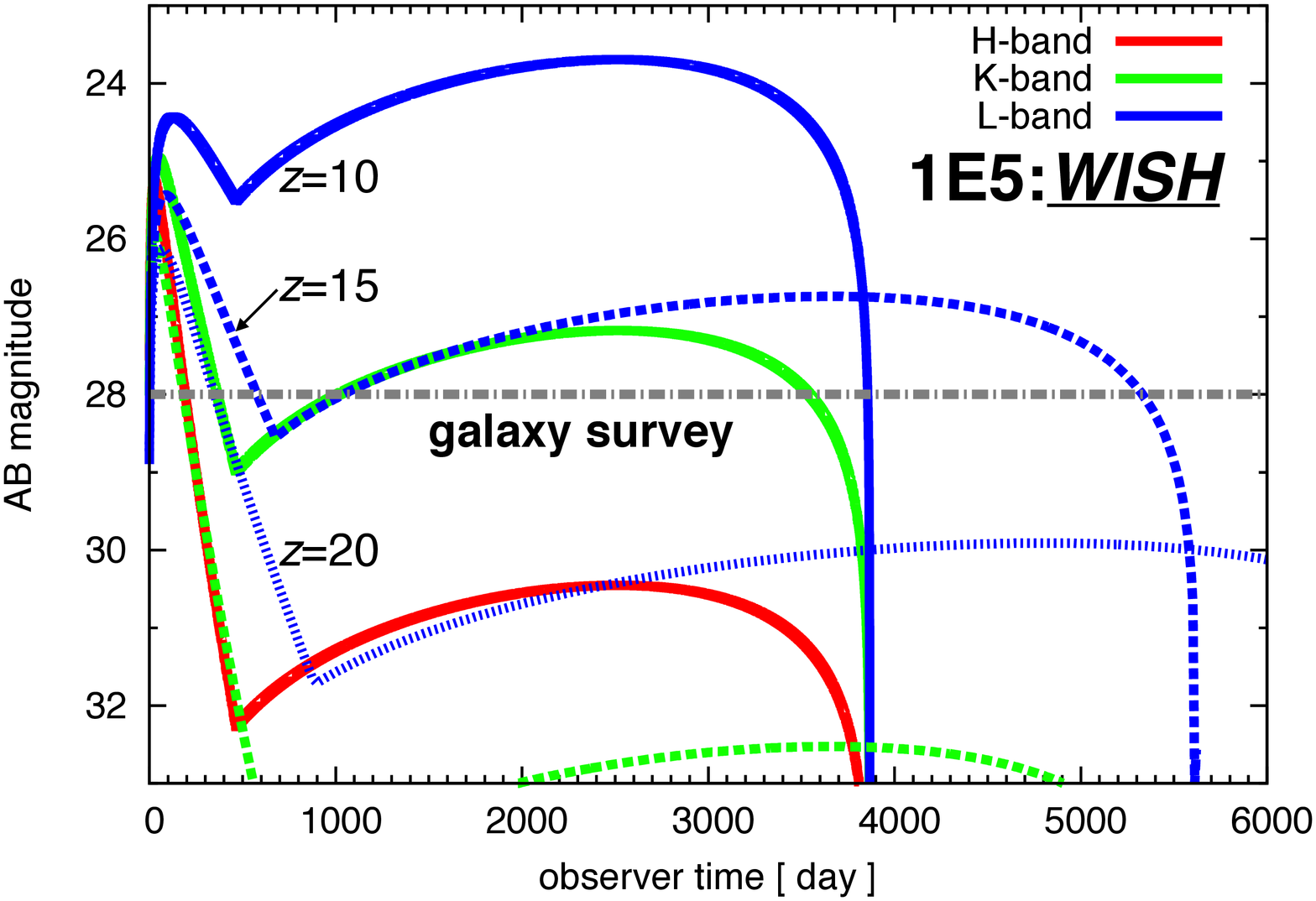}
\caption{Same as Fig. \ref{lc_1e5_euclid image}, but for the 1E5 SMSs observed with \textit{WISH}. 
We add the light curves in K- and L-bands with green and blue curves.
The solid, dashed and dotted curves correspond to the redshifts of the progenitors $z=10$, $15$, and $20$, respectively. 
Before the recombination sets in ($t_{\rm{obs}}\lesssim500\,\rm{days}$), the light curves show the bright first peaks.
After $t_{\rm{obs}}\gtrsim500\,\rm{days}$, they show the second peaks.}
\label{lc_1e5_wish image}
\end{figure}

\begin{figure}[!t]
\centering
\includegraphics[scale=0.3,angle=270]{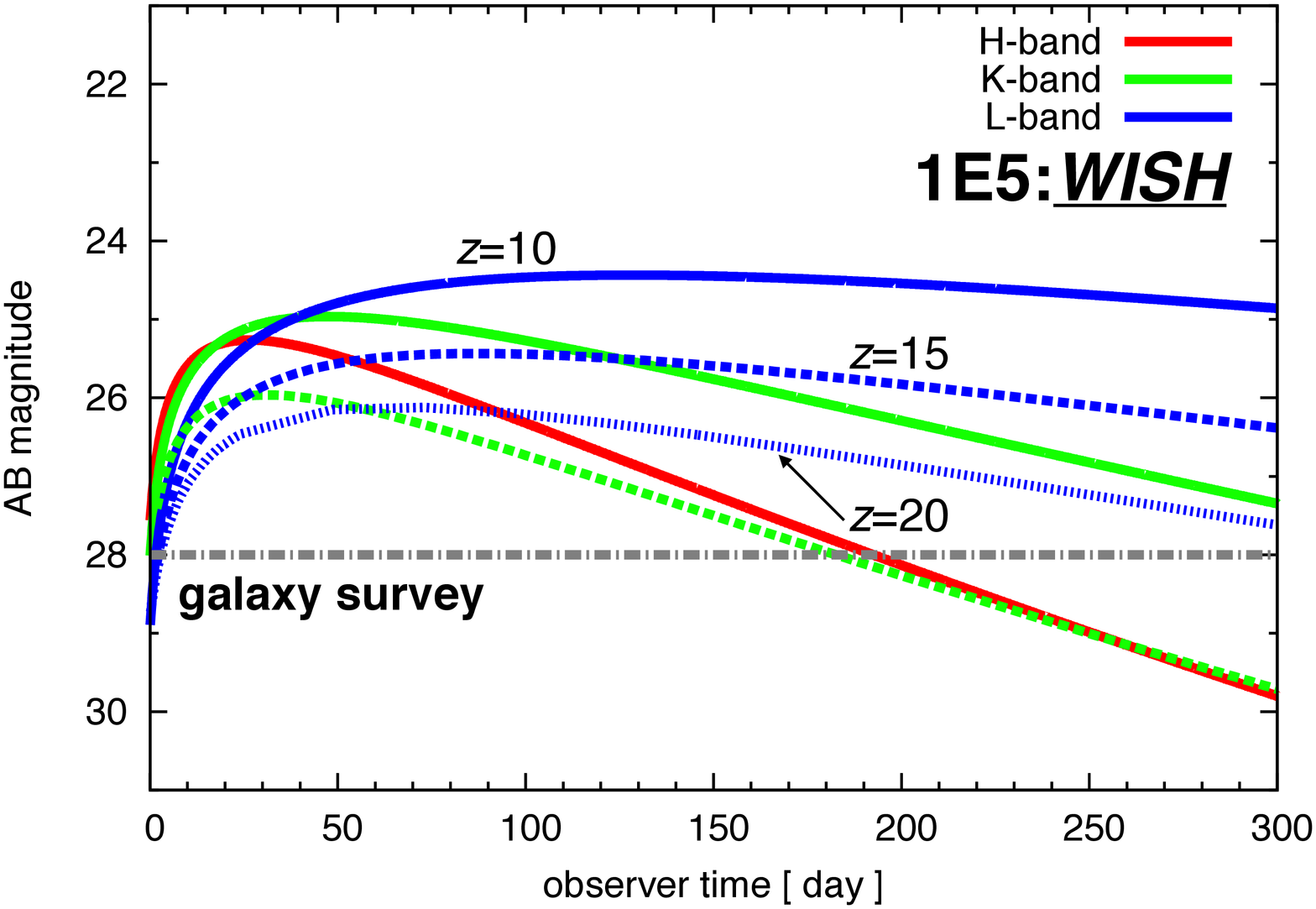}
\caption{Same as Fig. \ref{lc_1e5_wish image}, but we zoom in on the first $300$ days (rising first peak).}
\label{lc_1e5_wish2 image}
\end{figure}

In Figs. \ref{lc_1e5_euclid image} - \ref{lc_1e5_wish2 image}, we show the light curves of the cocoon emission for the 1E5 SMS model, and compare them with the detection limit of \textit{Euclid}, \textit{WFIRST}, and \textit{WISH}, respectively.
The horizontal axis represents the time since the cocoon fireballs start homologous expansion in the observer frame.
The vertical axis shows the observed AB magnitude.
\textit{Euclid} has a difficulty to detect the cocoon emission in galaxy survey because the cocoon emission is not bright enough.
In the SN survey, \textit{Euclid} can detect the first peak of the cocoon emission for $\sim80$ days.
\textit{WFIRST} can detect only the first peak of the cocoon emission at $z=10$.
In the galaxy survey, the first peak can be observed for $\sim90$ and $120$ days in H- and F184-bands, respectively, while in the SN survey, it can be observed for $\sim 300$ days.
It should be noted that \textit{WFIRST} uses only J- and H-bands in the SN survey (see Table \ref{telescope property table}).
At $z=15$, photons at the F184-band are strongly absorbed, so that the cocoon emission cannot be detected with \textit{WFIRST}.
\textit{WISH} has the deeper detection limits and more bands than those of \textit{Euclid} and \textit{WFIRST} in the galaxy survey.
At $z=10$, \textit{WISH} can detect the first peak in all bands with durations of longer than 100 days.
In L-band, \textit{WISH} will observe the whole of the light curves with the duration of $\sim3800\rm{\,days}$.
In K-band, the first and second peak are detected separately.
At $z=15$, \textit{WISH} will detect the first peak in K- and L-bands.
In L-band, the second peak is also detectable.
At $z=20$, only L-band can be used.
\textit{WISH} can detect the first peak with the duration of $\sim400\rm{\,days}$.

\begin{figure}[!t]
\centering
\includegraphics[scale=0.3,angle=270]{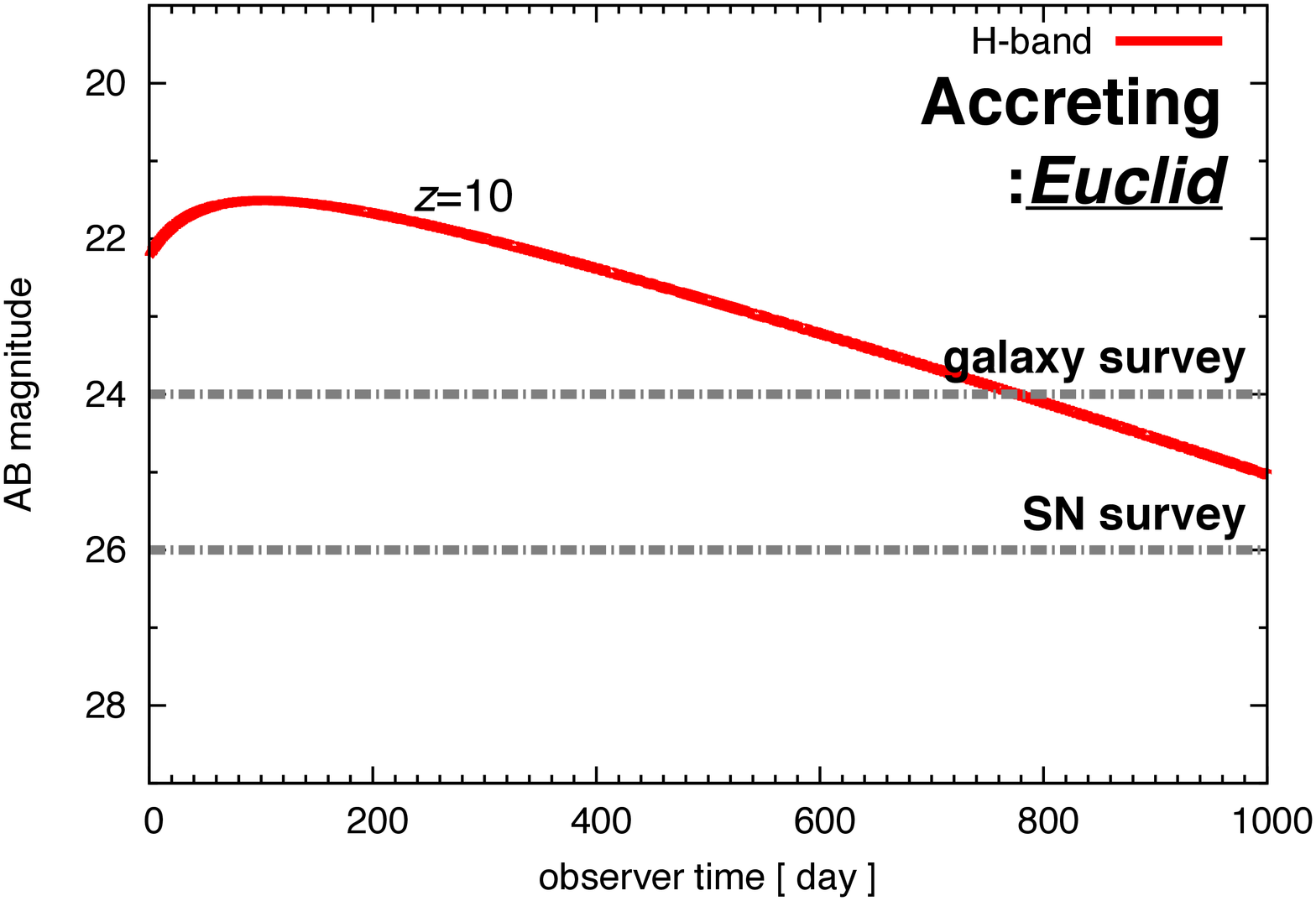}
\caption{Same as Fig. \ref{lc_1e5_euclid image}, but for the Accreting SMS model.}
\label{lc_acc_euclid image}
\end{figure}

\begin{figure}[!t]
\centering
\includegraphics[scale=0.3,angle=270]{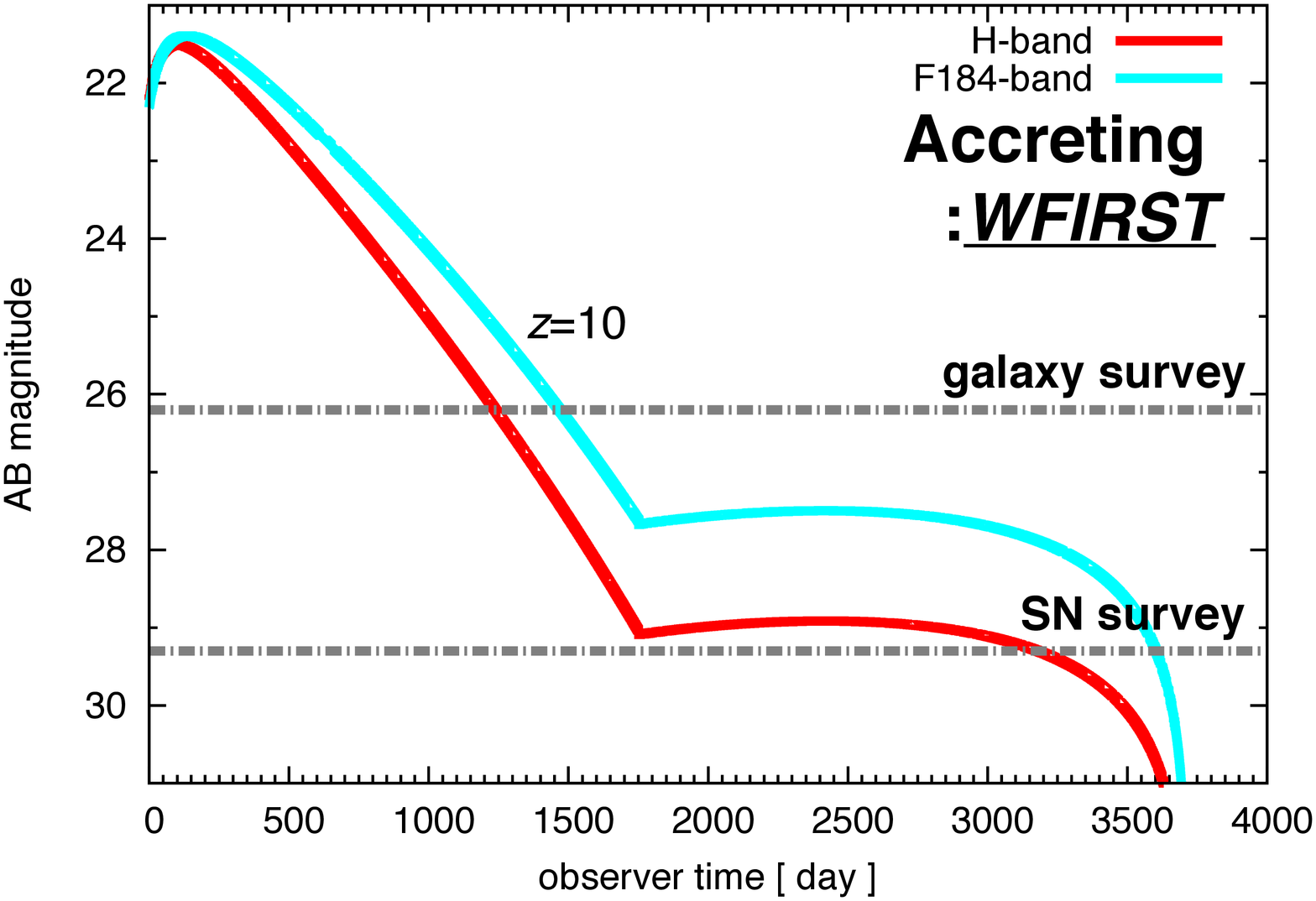}
\caption{Same as Fig. \ref{lc_1e5_wfirst image}, but for the Accreting SMS model.}
\label{lc_acc_wfirst image}
\end{figure}

\begin{figure}[!t]
\centering
\includegraphics[scale=0.3,angle=270]{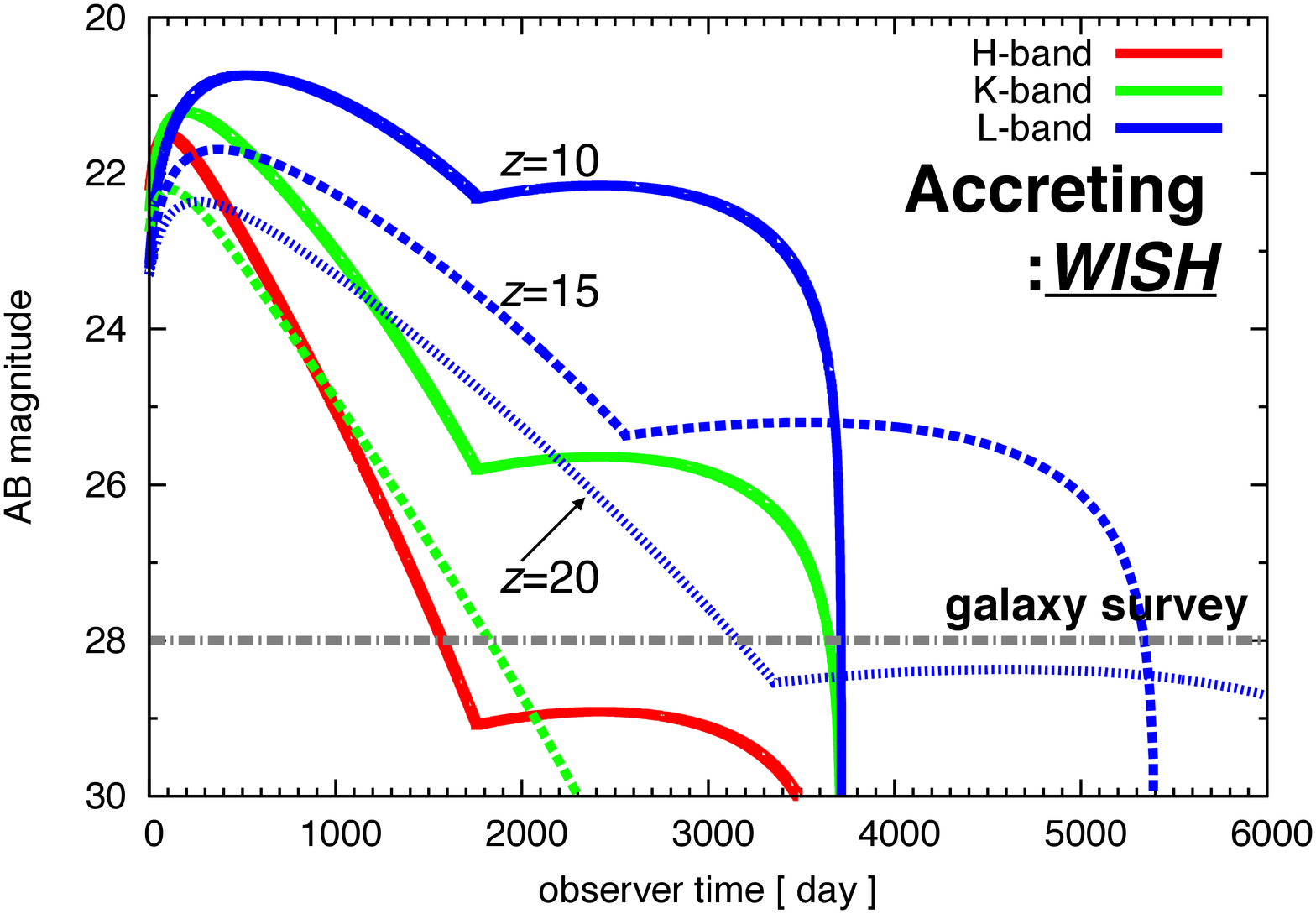}
\caption{Same as Fig. \ref{lc_1e5_wish image}, but for the Accreting SMS model.}
\label{lc_acc_wish image}
\end{figure}

In Figs. \ref{lc_acc_euclid image} - \ref{lc_acc_wish image}, we show the light curves of the cocoon emission obtained from the Accreting SMS model.
In the Accreting SMS model, the duration of the first peak is comparable to that of the second one, because the Accreting SMSs have larger radius than the 1E5 SMSs (Eq. \ref{recombination time}).
The large radius also makes cocoon emission about 10 times brighter than those of the 1E5 SMS model (Eq. \ref{cocoon luminosity1}).
\textit{Euclid} can detect the first peak in H-band for more than 600 days.
\textit{WFIRST} also detects the first peak in all available bands for $\gtrsim1000\rm{\,days}$.
In the SN survey, it may observe the second peak with the duration of $\sim3000\rm{\,days}$ in H-band.
\textit{WISH} can detect the first peak up to $z=20$ in all bands for $1000-3000\rm{\,days}$.
In particular, the first and second peaks are observed for $\sim3700-5400\rm{\,days}$ in K- ($z=10$) and L-bands ($z=10$ and $15$), respectively.

\section{Redshift Determination and Event Rate}\label{Event rate}
\subsection{Redshift Determination}
In the previous section, we find that the cocoon emission from SMSs may become ultra-luminous SNe. They can be detectable with the future wide-field NIR survey telescopes, such as \textit{Euclid}, \textit{WFIRST}, and \textit{WISH}.
In this section, we discuss the survey strategy of these ultra-luminous SNe.

The NIR survey telescopes have two survey modes, the SN and galaxy surveys.
In the SN survey, they visit the same area many times.
By comparing the images taken at different times, they can detect cocoon emissions as transient events.
As shown in Table \ref{telescope property table}, in the SN survey, the telescopes cover much narrower areas than those in the galaxy survey.
This will reduce the number of detectable events (see the next section).
However, the telescopes observe more deeply than they do in the galaxy survey.
The sensitive observation with high cadence in the SN survey will enable us to detect the cocoon emissions without confusing with other events.

In the galaxy survey, they survey large areas and take images of many galaxies.
They can detect cocoon emissions as one of the brightest stationary sources at high redshift.
Since the survey area in the galaxy survey is larger than that in the SN survey, we can expect larger event rate in this mode.
A color-color diagram is useful to discriminate the cocoon emission from other objects such as QSOs or brown dwarves \citep{2006ApJ...637...80M,2013MNRAS.435.2483T}.
Because of Lyman-$\alpha$ absorption in short wavelengths, we can select the candidates of the cocoon emission as red objects in a color-color diagram.
It should be noted that the cocoon emission is one of the brightest sources at high redshift, making it easy to disentangle the cocoon emissions from high-z galaxies.
We estimate the luminosity of the cocoon emissions in Eq. \eqref{cocoon luminosity1} as $L_{\rm{cocoon}}\sim10^{45}\,{\rm erg\,s^{-1}}$, which is about 100 times larger than that of ordinary galaxies $L_{\rm{gal}}\sim10^{10}L_{\odot}$.
We do not have the luminosity functions of galaxies at $z\gtrsim10$, but we know the luminosity function at $z\sim8$ obtained by BoRG survey as $\phi(L)=\phi_{*}(L/L_*)^{\alpha}\exp[{-L/L_*}]$, where $\phi_*\simeq4\times10^{-4}{\,\rm{Mpc^{-3}}}$, $\alpha\simeq-2$, and $L_*\simeq1.1\times10^{10}L_{\odot}$ \citep{2012ApJ...760..108B}.
By assuming that this function holds at $z\gtrsim8$, we estimate the number density of the galaxies, whose luminosity is comparable with that of the cocoon emissions, as $\sim10^{-8}{\,\rm{Gpc^{-3}}}$.
Therefore, the possibility that we detect bright galaxies as the cocoon emissions is extremely low (see the next section for the event rate of the cocoon emissions).
We can also use the time variability of cocoon emissions to disentangle cocoon emissions from galaxies.
Follow-up observations will find that the color of cocoon emissions becomes redder than that at the first detection (see below).

In Fig. \ref{cocoon_acc_z10_sed image}, we show the temporal evolution of the spectral energy distribution (SED) of the cocoon emission at $z=10$.
It is obtained from the Accreting SMS model.
The horizontal axis corresponds to the wavelength in the observer frame.
The vertical axis represents the observed flux density.
The dark-grey shaded region shows the wavelength region in which photons are absorbed by neutral hydrogen in the intergalactic medium (IGM). 
The light-grey shaded regions correspond to the J-, H-, K-, and L-bands from left to right, respectively.
The horizontal dashed lines represent the best sensitivities of \textit{Euclid}, \textit{WFIRST}, and \textit{WISH}.
We also show the sensitivities of \textit{James Webb Space Telescope} (\textit{JWST}) \citep{2006SSRv..123..485G}, in the spectroscopic and photometric observations with the dash-dotted and dotted curves, respectively.
\textit{JWST} is useful for the follow-up spectroscopy.

\begin{figure}[!t]
\centering
\includegraphics[scale=0.3,angle=270]{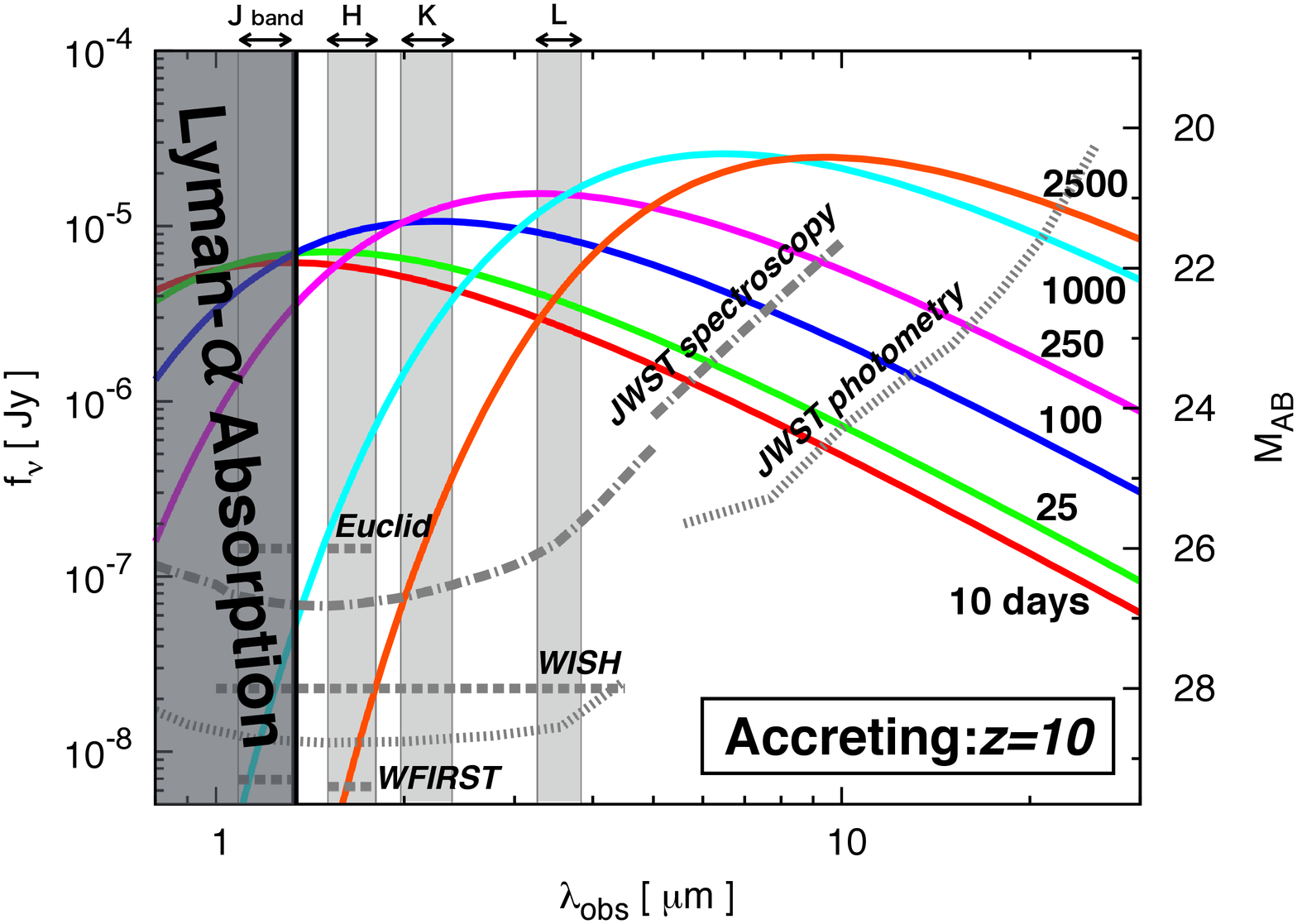}
\caption{Time evolution of the spectral energy distribution (SED) of the cocoon emission from the Accreting SMS at $z=10$. The horizontal axis shows the wavelength in the observer frame. The vertical axis represents the flux density. The SEDs after 10, 25, 100, 250, 1000, and 2500 days are shown with the red, green, blue, magenta, light-blue, and orange solid curves, respectively. We represent the wavelength range in which photons suffer from Lyman-$\alpha$ absorption with dark-shaded region. We also show the regions which correspond to J-, H-, K- and, L-bands with light-shaded regions. The grey dashed lines in J-, H-, K-, and L-bands represent the best sensitivities of \textit{Euclid}, \textit{WFIRST}, and \textit{WISH}. The grey dash-dotted and dotted curves also show the sensitivities of \textit{JWST} in the spectroscopic and photometric observations\footnote{http://www.stsci.edu/jwst/science/sensitivity}, respectively. The Lyman-$\alpha$ damping at $\lambda_{\rm{obs}}\lesssim 1.34[(1+z)/11]\,\mu\rm{m}$ tells us the redshift of the burst.}
\label{cocoon_acc_z10_sed image}
\end{figure}

From Fig. \ref{cocoon_acc_z10_sed image}, we find that the flux is above the detection limit of \textit{JWST} spectroscopy for $t_{\rm{obs}}\lesssim 1000\,\rm{days}$.
In this case, we can identify the absorption edge of the SED, which is made by Lyman-$\alpha$ absorption.
Then, the distance or redshift of the event may be measured spectroscopically by the Gunn-Peterson trough, as is often done in the QSO observations \citep{1965ApJ...142.1633G}.

We can obtain the spectroscopic information around the first peak of the cocoon emission for $t_{\rm{obs}}\lesssim 2500\,\rm{days}$.
If the SED can be taken around of the first peak, we can estimate the bolometric luminosity at the first peak $L_{\rm{1st}}$.
As shown in Eq. \eqref{cocoon luminosity1}, the bolometric luminosity decreases monotonically, and we use Eq. \eqref{cocoon luminosity1} at $t=0$ for the luminosity $L_{\rm{1st}}$ below.
The decreasing rate of the bolometric luminosity gives the diffusion time of the cocoon fireball  in the observer frame $t_{\rm{d}}(1+z)$.
We will also be able to estimate the photospheric velocity $v_{\rm{ph}}$ from e.g., the P Cygni profile of the hydrogen Balmer line~($\lambda_{{\rm H}\alpha} = 0.656(1+z) \mu \rm m$).
It should be noted that around the first peak, the photospheric velocity $v_{\rm{ph}}$ is evaluated by the cocoon velocity $\vc$ (Eq. \ref{photospheric velocity}).
The observables in the first peak are useful to estimate the progenitor's parameters, because the first peak is brighter in band ranges of the \textit{Euclid}, \textit{WFIRST}, and \textit{WISH} photometry and the \textit{JWST} spectroscopy.

From Eqs. \eqref{diffusion time}, \eqref{cocoon luminosity1}, and \eqref{photospheric velocity}, the energy $\Ec$, mass $\Mc$, and initial radius $\Rc$ of the cocoon fireball can be obtained as functions of the observables, $L_{\rm{1st}}$, $t_{\rm{d}}$, and $v_{\rm{ph}}$, as
\begin{eqnarray}
\Mc&=&\frac{4\pi{c}}{3\kappa}t_{\rm{d}}^{2}v_{\rm{ph}}\label{a}\\
&\simeq&1.8\times10^{3}\,\Msun\ t_{\rm{d,7}}^2 v_{\rm{ph, 10}}, \nonumber\\
\Ec&=&\frac{4\pi{c}}{5\kappa}t_{\rm{d}}^{2}v_{\rm{ph}}^3\label{b}\\
&\simeq&2.2\times10^{55}\,{\rm{erg}}\ t_{\rm{d,7}}^2 v_{\rm{ph, 10}}^3, \nonumber\\
\Rc&=&\frac{5\kappa}{2\pi{c}}L_{\rm{1st}}v_{\rm{ph}}^{-2}\label{c}\\
&\simeq&9.3\times10^{13}{\rm{\,cm}}\ L_{\rm{1st, 45}} v_{\rm{ph, 10}}^{-2}, \nonumber
\end{eqnarray}
where $t_{\rm{d,7}}=t_{\rm{d}}/{10^{7}{\rm{\,s}}}$, $L_{\rm{1st, 45}} = L_{\rm{1st}}/{10^{45}{\rm{\,erg\,s^{-1}}}}$, and $v_{\rm{ph, 10}} = v_{\rm{ph}}/10^{10}{\rm{\,cm\,s^{-1}}}$.

When the bolometric luminosity does not change so much around the first peak, we can use the total duration of the cocoon emission $\Delta{t_{\rm{co}}}$ in order to estimate the diffusion time $t_{\rm{d}}$.
This situation actually occurs when the recombination starts at much faster than the diffusion timescale.
From Eqs.  \eqref{c}, \eqref{recombination time}, and \eqref{plateau duration}, we obtain the diffusion timescale as
\begin{eqnarray}
t_{\rm{d}}&=&\frac{(4\pi\sigma_{\rm{SB}}T_{\rm{ion}}^{4})^{1/5}}{7^{1/2}}L_{\rm{1st}}^{-1/5}v_{\rm{ph}}^{2/5}\Delta{t}_{\rm{co}}^{7/5}\\
&\simeq&5.9\times10^{6}{\rm{\,s}}\ L_{\rm{1st,45}}^{-1/5}v_{\rm{ph,10}}^{2/5}\Delta{t}_{\rm{co,7}}^{7/5},\nonumber
\end{eqnarray}
where $\sigma_{\rm{SB}}$ is the Stefan-Boltzmann constant and $\Delta{t_{\rm{co,7}}} = \Delta{t_{\rm{co}}}/{10^{7}{\rm{\,s}}}$.
We can substitute this expression for $t_{\rm{d}}$ in Eqs. \eqref{a} and \eqref{b}.

\cite{1985SvAL...11..145L} and \cite{1993ApJ...414..712P} also gave the expressions for the parameters of Type \2P SNe progenitors, by using the duration, luminosity, and photospheric velocity in the plateau phase (i.e., the second peak).
Their expressions have been used to derive the explosion parameters of Type \2P SNe \citep{2003ApJ...582..905H,2013MNRAS.433.1871B,2015arXiv150901721D}.
When we use the observables of the second peak, i.e., the bolometric luminosity $L_{\rm{2nd}}$ and the photospheric velocity $v_{\rm{ph}}(t_{\rm{2nd}})$, rather than $L_{\rm{1st}}$ and $v_{\rm{ph}}$, Eqs. \eqref{plateau duration}, \eqref{peak luminosity2}, and \eqref{photospheric velocity2} become dependent and we cannot solve for the parameters $\Mc$, $\Ec$, and $\Rc(0)$.
This is not pointed out in the previous studies \citep{1985SvAL...11..145L,1993ApJ...414..712P}, in which they use $\vc=(5\Ec/3\Mc)^{1/2}$ rather than $v_{\rm{ph}}(t_{\rm{2nd}})$, but we think that $v_{\rm{ph}}(t_{\rm{2nd}})$ (not $\vc$) is the observable quantity for the second peak.

From Eq. \eqref{cocoon mass2}, the progenitor mass can be estimated by using Eq. \eqref{a}
\begin{align}
M_{\ast} &\sim 6.4\times10^4\Msun\,\ {t_{\rm{d,7}}}^{2}v_{\rm{ph, 10}} \notag \\
& \times \biggl(\frac{\eta_{\rm{j}}}{6.2\times10^{-4}}\biggl)^{1/2}\biggl(\frac{\theta}{5^\circ}\biggl)^{-2},
\end{align}
If this is larger than the theoretical mass of any Pop \3 stars via hydrogen molecular cooling \citep{2014ApJ...781...60H}, the observation gives the first direct evidence of a SMS.

The spectroscopic observation also tells us the existence of metals in the cocoon fireballs \citep[see also, ][for probing the metal enrichment in the early IGM with GRB afterglows]{2012ApJ...760...27W}.
With taking the ratio of line strengths, we can infer the abundance of heavy elements.
If SMSs are really made of the primordial gas and the cocoon fireballs are metal free, we will observe only hydrogen and helium lines.
This is a direct evidence of Pop \3 stars.
Thus, the cocoon emission from SMSs provides us an unique opportunity to explore the Pop \3 SMSs.

\subsection{Event Rate}\label{event rate}
Here, we discuss the event rate of the cocoon emission.
The cumulative number of cocoon emission events observed with a telescope with a survey area $\Omega_{\rm{obs}}$ is calculated by
\begin{eqnarray}
\Delta{N(z)}=\int_0^{z}\Psi_{\rm{burst}}(z^\prime)4\pi{c}r(z^\prime)^2\biggl|\frac{dt}{dz}\biggl|dz^\prime\Delta{t}_{\rm{obs}}\frac{\Omega_{\rm{obs}}}{4\pi},
\label{event rate}
\end{eqnarray}
where $\Psi_{\rm{burst}}(z)$, $r(z)$ and $\Delta{t}_{\rm{obs}}$ are the intrinsic burst rate in the comoving volume, the comoving distance, and the observation time, respectively.
The intrinsic event rate $\Psi_{\rm{burst}}(z)$ is related to the SMS or DCBH formation rates ($\frac{dn_{\rm{SMS}}}{dt}$ or $\frac{dn_{\rm{DCBH}}}{dt}$), which are studied in previous studies.
In the following rate estimation, we simply assume that the SMS GRB event rate $\Psi_{\rm{burst}}$ is proportional to the SMS formation rate $\frac{dn_{\rm{SMS}}}{dt}$ as, $\Psi_{\rm{burst}}=f_{\rm{GRB}}\frac{dn_{\rm{SMS}}}{dt}$, where $f_{\rm{GRB}}$ is a fraction of the GRB formation.
It should be noted that since the cocoon emission emits photons isotropically, the event rate is not suppressed by the jet beaming factor $\sim\theta^2$ like GRBs.

The formation theories of SMSs or DCBHs have a lot of unknown parameters, e.g, the star formation efficiency, the escape fraction of FUV photons from the host halo, the metal enrichment via galactic outflow, and the reionization history of the Universe.
We should determine which model predicts a correct formation rate by observations.

\begin{figure}[!t]
\centering
\includegraphics[scale=0.3,angle=270]{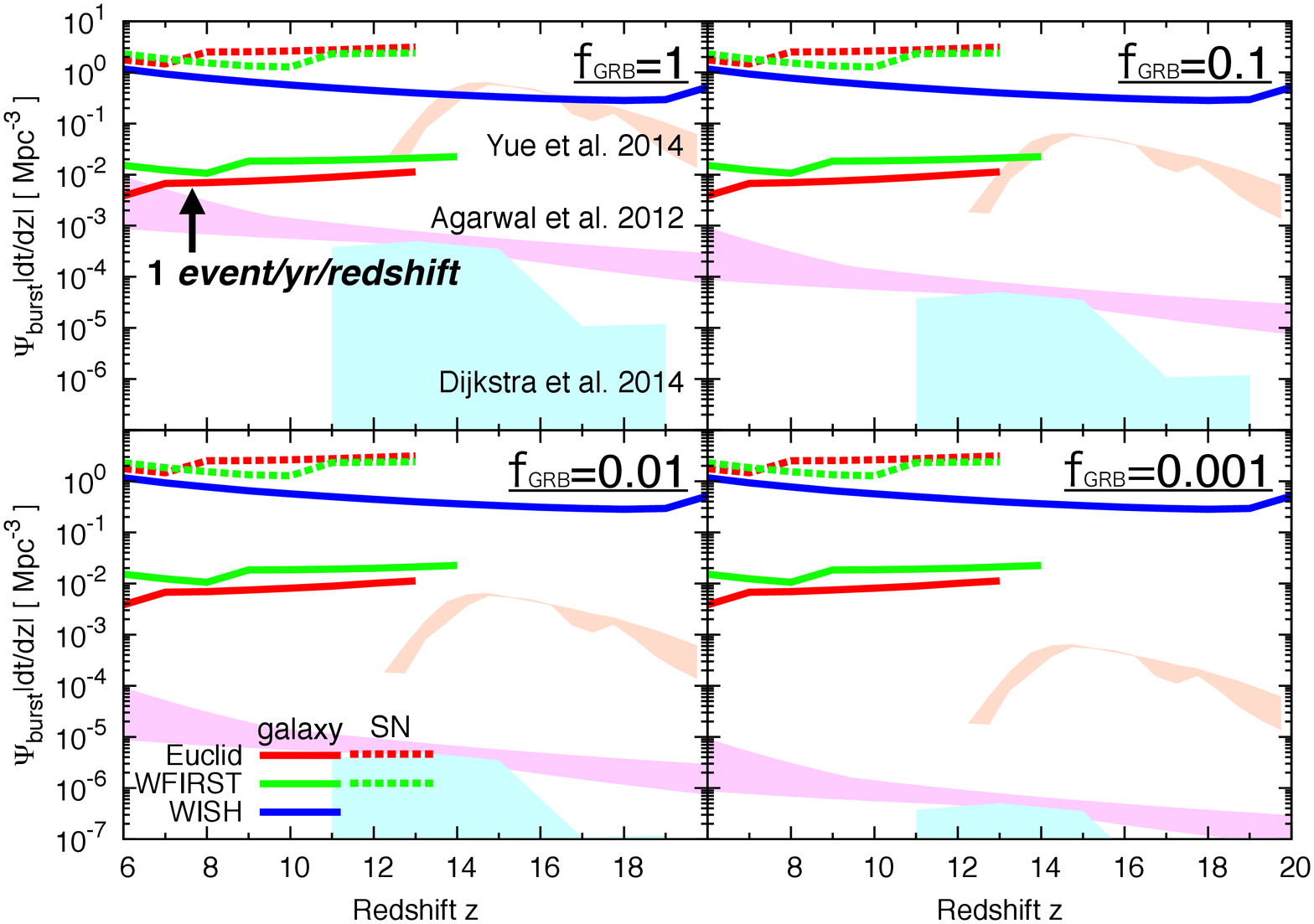}
\caption{SMS GRB event rates in comoving density per redshift converted from the SMS formation rates as $\Psi_{\rm{burst}}=f_{\rm{GRB}}\frac{dn_{\rm{SMS}}}{dt}$.
The pink, light-blue, and orange shaded regions represent the GRB rates given by the SMS formation rates which are calculated by \cite{2012MNRAS.425.2854A}, \cite{2014MNRAS.442.2036D} and \cite{2014MNRAS.440.1263Y}, respectively.
The upper-left, -right, bottom-left, and -right panels show the cases of the GRB fraction $f_{\rm{GRB}}=1$, $0.1$, $0.01$, and $0.001$, respectively.
The red, blue, and green curves show the formation rates which give 1 event $\rm{yr^{-1}}$ per redshift by observation with future telescopes \textit{Euclid}, \textit{WFIRST}, and \textit{WISH}, respectively. The solid and dashed curves represent the galaxy and SN survey modes, respectively.
\if{Formation rate of SMSs or DCBHs in comoving density per redshift. The pink, light-blue, and orange shaded regions represent the formation rates of SMSs or DCBHs calculated by \cite{2012MNRAS.425.2854A}, \cite{2014MNRAS.442.2036D} and \cite{2014MNRAS.440.1263Y}, respectively. The red, blue, and green curves show the formation rates which give 1 event $\rm{yr^{-1}}$ per redshift by observation with future telescopes \textit{Euclid}, \textit{WFIRST}, and \textit{WISH}, respectively. The solid and dashed curves represent the galaxy and SN survey modes, respectively.
We assume the galaxy survey mode in the observations of these telescopes and the Accreting SMSs in the calculations of the cocoon emission.}\fi
}
\label{event rate image}
\end{figure}

In Fig. \ref{event rate image}, we show the SMS GRB rates converted from the SMS formation rates, as $\Psi_{\rm{burst}}=f_{\rm{GRB}}\frac{dn_{\rm{SMS}}}{dt}$.
The vertical axis shows the SMS GRB rate in comoving volume per redshift.
Pink, light-blue, and orange shaded regions show the SMS GRB rates given by the SMS formation rates which are calculated by  \cite{2012MNRAS.425.2854A}, \cite{2014MNRAS.442.2036D}, and \cite{2014MNRAS.440.1263Y}\footnote{\cite{2014MNRAS.442.2036D} and \cite{2014MNRAS.440.1263Y} show only the cumulative number densities of DCBHs.
Then, we convert their results into the formation rates.}, respectively.
Since there are no constraints on the conversion parameter $f_{\rm{GRB}}$, we also consider four cases of the parameter values of $f_{\rm{GRB}}=1$, $0.1$, $0.01$, and $0.001$ and plot each case separately.
The case of $f_{\rm{GRB}}=1$ means that all SMSs produce ULGRBs and cocoon emissions.
This will be an upper limit of the SMS GRB rate and the true GRB rate will be lower than that.
In the local Universe, the GRB rate is about $10^{-3}$ times of the core-collapse SN rate \citep{2010MNRAS.406.1944W}.
The local fraction corresponds to $f_{\rm{GRB}}=0.001$.

We also show the formation rates which give event rates $=1\rm{\,event\,yr^{-1}}$ per redshift by the observations with \textit{Euclid}, \textit{WFIRST}, and \textit{WISH}, using red, green, and blue curves, respectively.
The solid and dashed curves represent the galaxy and SN survey modes, respectively.
These rates are obtained by equating the integrand in Eq. (\ref{event rate}) as unity.
Since the probability of detections increases when the duration of events $\Delta{t_{\rm{co}}}$ is longer than the observation time $\Delta{t_{\rm{obs}}}$, we multiply Eq. (\ref{event rate}) by the modifying factor $\Delta{t_{\rm{co}}}/\Delta{t_{\rm{obs}}}$.

\if{
{\bf In Fig. \ref{event rate image}, we equate the intrinsic burst rate $\Psi_{\rm{burst}}(z)$ to the SMS formation rate, assuming that all SMSs could produce ULGRBs and cocoon emissions.
Practically, the fraction of SMSs that could produce GRBs is uncertain.
In the local Universe, the GRB rate is only $\sim10^{-3}$ times the core-collapse SN rate \citep{2010MNRAS.406.1944W}.
Such a small rate could be explained partly because the GRB progenitors should rotate at nearly the breakup velocity \citep{2005A&A...443..643Y,2006ApJ...637..914W}.
It is possible, however, that all SMSs can acquire enough angular momentum to rotate at nearly the breakup velocity for producing ULGRBs.
\cite{2014MNRAS.445L.109I} numerically studied the gravitational collapse of warm gas clouds in atomic cooling halos and showed that the inflowing gas onto the cloud center has a rotational velocity whose value is about half the Keplerian rotation velocity $v_{\rm{rot}}\sim{v_{\rm{K}}}/2$.
When the inflow is accreted onto the central protostar, it supplies the protostar with angular momentum.
Moreover, the accretion could continue over the lifetime of a SMS.
\cite{2013ApJ...778..178H} studied the evolution of SMSs with rapid mass accretion, and found that SMSs could have very bloated envelopes and their effective temperatures could be too low ($\sim 5000$ K) to trigger UV feedback onto the accretion flow.
Therefore, the angular momentum supply to a SMS could be kept over the lifetime and the SMS could rotate at about the breakup velocity at the precollapse phase.
According to the above consideration, we assume that the event rate of SMS GRBs is comparable to the formation rate of SMSs.}
}\fi

If a model predicts more formation rate than the rates represented by solid or dashed curves in Fig. \ref{event rate image}, we expect that we observe the cocoon emission more than $1\,\rm{event\ yr^{-1}}$.
From Fig. \ref{event rate image}, we see that the models studied by \cite{2014MNRAS.440.1263Y} predict much more event rate than those studied by \cite{2012MNRAS.425.2854A} and \cite{2014MNRAS.442.2036D}.
Actually, the cumulative event rate reaches $\sim30\ (f_{\rm{GRB}}/1)\ \rm{events\ yr^{-1}}$ for the model studied by  \cite{2014MNRAS.440.1263Y}.
On the other hand,  \cite{2012MNRAS.425.2854A} and \cite{2014MNRAS.442.2036D}'s models predict $\lesssim1\ (f_{\rm{GRB}}/1)\,\rm{event\ yr^{-1}}$.\footnote{The formation rate of DCBHs depends strongly on the critical intensity of the FUV field which is needed for the SMS formation.
\cite{2012MNRAS.425.2854A} and \cite{2014MNRAS.442.2036D} calculated the formation rate by adopting $J_{\rm{crit}}=30-300$, where $J_{\rm{crit}}$ is the intensity at $h\nu=12.4\,\rm{eV}$ and in units of $10^{-21}\,\rm{erg\,s^{-1}\,cm^{-2}\,Hz^{-1}\,sr^{-1}}$.
Recently, \cite{2014MNRAS.445..544S} obtained $J_{\rm{crit}}\simeq1000$ by considering the realistic spectra of the FUV field from metal-poor galaxies.
This value is larger than those used in \cite{2012MNRAS.425.2854A} and \cite{2014MNRAS.442.2036D}, so that they may overestimate the formation rate to some extent \citep[see also,][]{2015MNRAS.450.4350I}.}
Thus, by the actual observations with the future telescopes, we can select preferred models.

For its 11 yrs of operation, BAT onboard \textit{Swift} satellite has never detected ULGRBs from SMSs.
However, this result gives little constraint on the SMS GRB event rate nor DCBH formation rate.
The detectability of SMS GRBs with BAT depends on the models of a progenitor star and the prompt emission \citep{2015ApJ...810...64M}.
Then, for some models, BAT does not have enough sensitivity to detect the GRBs.
On the other hand, the cocoon emissions are detectable with all future telescopes, and their simple emission mechanism gives few uncertainties to the detectability.
We can discuss the DCBH formation rate, which is essentially unconstrained, more robustly using the cocoon emission than using the SMS GRB.

\section*{acknowledgments}
We thank K. Inayoshi, K. Kashiyama, K. Sugimura, and H. Yajima for fruitful discussion.
We also thank for A. Heger for giving us his numerical data which were used in the previous paper, which was written with him, and the present paper.
This work is supported in part by the Grant-in-Aid from the
Ministry of Education, Culture, Sports, Science and Technology (MEXT)
of Japan, Nos. 261051 (DN) 24103006, 26287051, 24000004, 26247042 (KI)
24103006, 15H02087 (TN).

\appendix

\section{A. Analytical estimate for cocoon parameters}\label{Analytical estimate for cocoon parameters}
It is very useful to develop analytical formulae for cocoon parameters in order to see the dependences on progenitors' properties.
We estimate the cocoon energy $\Ec$ and mass $\Mc$ at the jet breakout analytically in this appendix.
For more details of our jet propagation model, see \cite{2015ApJ...810...64M}.

First, we calculate the cocoon energy $\Ec$ at the jet breakout defined as the energy stored into a cocoon component during the jet propagation in a progenitor as 
\begin{eqnarray}
\Ec=\eta_{\rm{c}}\int^{t_{\rm{b}}}_{t_{\rm{in}}}L_{\rm{j}}dt,
\label{cocoon energy1}
\end{eqnarray}
where $\eta_{\rm{c}}$, $t_{\rm{in}}$, $t_{\rm{b}}$ and $L_{\rm{j}}$ are the fraction of matter flowing into the cocoon from the jet head, the jet injection time, the jet breakout time, and the jet luminosity, respectively.
It should be noted that we set the origin of time when the progenitor starts to collapse.
We evaluate the efficiency parameter $\eta_{\rm{c}}$ as unity because the jet head is sub relativistic in the progenitor.
In the following, we estimate the quantities $t_{\rm{in}}$, $t_{\rm{b}}$, and $L_{\rm{j}}$.

Let us first estimate the jet luminosity $L_{\rm{j}}$.
In our study, we consider the MHD mechanism as the jet formation process \citep{1977MNRAS.179..433B}.
In this mechanism, the jet luminosity $L_{\rm{j}}$ is given by the mass accretion rate onto the central BH as $L_{\rm{j}}=\eta_{\rm{j}}\dot{M}c^2$ \citep{2010MNRAS.402L..25K}, where the efficiency parameter $\eta_{\rm{j}}=6.2\times10^{-4}$ is calibrated to reproduce observed total energy of a jet when we apply our jet propagation prescription to Wolf-Rayet stars \citep{2011ApJ...726..107S}.
We assume that a mass shell at mass coordinate $M_{r}=\int_0^r4\pi{r}^2\rho{dr}$ falls onto the BH in its free-fall time defined as $t_{\rm{ff}}(r)=\sqrt{\pi^2r^3/8GM_r}$, where $G$ is the gravitational constant.
Then we obtain the mass accretion rate as $\dot{M}=(dM_r/dr)/(dt_{\rm{ff}}/dr)$. 

While the launching mechanisms of relativistic jets are still uncertain, the following two mechanisms are often discussed: (i) MHD mechanism \citep{1977MNRAS.179..433B} and (ii) neutrino and antineutrino annihilation mechanism \citep{1999ApJ...518..356P}.
In the latter model, as the mass of a central BH gets larger, the energy density in the accretion disk becomes smaller, so that the jet could be quenched before the jet breakout  \citep{2011MNRAS.410.2302Z,2011ApJ...726..107S}.
On the other hand, in the MHD jet model, the jet luminosity depends only on the mass accretion rate and is proportional to the accretion rate, $L_{\rm{j}}\propto\dot{M}$.
Therefore, as long as the mass accretion rate onto the central BH is large and there is a global magnetic field, a powerful jet could be sustained regardless of the BH mass nor the energy density around the BH.
In our previous paper, we find that in the SMS case, high mass accretion rates ($\dot{M}\gtrsim0.1\,\Msun/\rm{s}$) could last for more than $10^4\rm{\,s}$, which is longer than the jet breakout time $\sim 4000\rm{\,s}$ \citep[see Fig. 3 in][]{2015ApJ...810...64M}.

\begin{table}[!t]
\begin{center}
\caption{Parameters of density profiles}
\label{density profile table}
\begin{tabular}{ccc}
\tableline\tableline
Progenitor Model & 1E5 & Accreting \\
\tableline
$\rho_{\rm{core}}$ [$\rm{g\,cm^{-3}}$] & $4.6\times10^{6}$ & $2.3$\\
$R_{\rm{core}}$ [cm] & $1.0\times10^{10}$ & $1.0\times10^{12}$\\
$R_*$ [cm] & $5.8\times10^{13}$ & $1.4\times10^{15}$\\
$M_{\rm{core}}$ [$\Msun$] & $9.6\times10^{3}$ & $4.8\times10^{3}$\\
\tableline
\end{tabular}
\end{center}
\end{table}

We see that once we know the density profile $\rho(r)$, we can calculate the quantities $M_r$, $t_{\rm{ff}}$, and $\dot{M}$.
SMSs have a density profile proportional to $r^{-3}$ at their radiation-pressure-dominated envelope \citep{1999ApJ...510..379M}.
Then, we approximate the density profile as follows,
\begin{eqnarray}
\rho(r)=\begin{cases}
\rho_{\rm{core}}&\text{$r<R_{\rm{core}}$,}\\
\rho_{\rm{core}}(\frac{r}{R_{\rm{core}}})^{-3}&\text{$R_{\rm{core}}<r<R_*$,}
\end{cases}
\label{density profile}
\end{eqnarray}
where $\rho_{\rm{core}}$ and $R_{\rm{core}}$ are the density and radius of the stellar core, where the density is constant.
In Table \ref{density profile table}, we show these parameters for our progenitor models \citep[these density profiles are shown in Figs. 1 and 6 in][]{2015ApJ...810...64M}.
With this simple density profile (\ref{density profile}), we calculate the mass coordinate $M_r$, free-fall time $t_{\rm{ff}}$, and their derivatives as follows,
\begin{eqnarray}
M_r&=&M_{\rm{core}}+3M_{\rm{core}}\ln\biggl(\frac{r}{R_{\rm{core}}}\biggl)\,\,\,\,\,\,(R_{\rm{core}}<r),
\label{mass coordinate}\\
\frac{dM_r}{dr}&=&\frac{3M_{\rm{core}}}{r},
\label{mass coordinate derivative}\\
\frac{dt_{\rm{ff}}}{dr}&=&\frac{3}{2}\bigg(\frac{\pi^2}{8G}\biggl)^{1/2}\biggl(\frac{r}{M_r}\biggl)^{1/2}\biggl[1-\frac{M_{\rm{core}}}{M_{r}}\biggl]\simeq\frac{3}{2}\biggl(\frac{\pi^2}{8G}\biggl)^{1/2}\biggl(\frac{r}{M_r}\biggl)^{1/2}.
\end{eqnarray}
Then, we obtain the mass accretion rate as
\begin{eqnarray}
\dot{M}\simeq\frac{3M_{\rm{core}}}{r}\frac{2}{3}\biggl(\frac{8G}{\pi^2}\biggl)^{1/2}\biggl(\frac{M_r}{r}\biggl)^{1/2}\simeq\frac{2M_{\rm{core}}}{t}.
\label{mass accretion rate}
\end{eqnarray}

The jet injection time may depend on the detail of the jet-launching mechanism.
However, we have showed that whenever the jet is launched from the central BH and accretion disk system, as long as the injection time is much shorter than the jet breakout time, the result does not change so much \citep{2015ApJ...810...64M}.
For simplicity, we consider the free-fall time of the stellar core as the jet injection time, 
\begin{eqnarray}
t_{\rm{in}}=\sqrt{\frac{\pi^2R_{\rm{core}}^3}{8G{M_{\rm{core}}}}}\simeq9.7\times10^{-1}\,{\rm{s}}\,\biggl(\frac{\rho_{\rm{core}}}{4.6\times10^6\,\rm{g\,cm^{-3}}}\biggl)^{-1/2}.
\label{jet injection time}
\end{eqnarray}

The jet breakout time is estimated by $t_{\rm{b}}\simeq{R_*}/\beta_{\rm{h}}c$, where $\beta_{\rm{h}}$ is the jet head velocity divided by the speed of light $c$. 
From the conservation of the momentum and energy flux at the jet head, the velocity $\beta_{\rm{h}}$ is given by the jet luminosity and the stellar density as \citep{2003MNRAS.345..575M,2011ApJ...740..100B}
\begin{eqnarray}
\beta_{\rm{h}}&\simeq&\biggl(\frac{L_{\rm{j}}}{\rho(r_{\rm{h}})c^3\Sigma_{\rm{h}}}\biggl)^{1/2}\simeq\biggl(\frac{8\eta_{\rm{j}}}{3c\theta^2}\biggl)^{1/2}\biggl(\frac{r_{\rm{h}}}{t}\biggl)^{1/2}\nonumber\\
&\simeq&\frac{8\eta_{\rm{j}}}{3\theta^2}\simeq2.2\times10^{-1}\,\biggl(\frac{\eta_{\rm{j}}}{6.2\times10^{-4}}\biggl)\biggl(\frac{\theta}{5^\circ}\biggl)^{-2}.
\label{jet head velocity}
\end{eqnarray}
In the first line, the quantity $\Sigma_{\rm{h}}=\pi(r_{\rm{h}}\theta)^2$ means the cross section of the jet head.
From the first line to the second line, we use the fact that the jet head position $r_{\rm{h}}$ is given by $r_{\rm{h}}\simeq{\beta_{\rm{h}}c}t$.
As shown in Eq. (\ref{jet head velocity}), the jet head velocity is constant in the density profile of $\rho\propto{r^{-3}}$ 
for a constant opening angle $\theta$ \citep{2015ApJ...810...64M}.
Then, we obtain the jet breakout time as 
\begin{eqnarray}
t_{\rm{b}}\simeq
\frac{3\theta^2 R_{*}}{8\eta_{\rm{j}}c}\simeq
8.9\times10^{3}\,{\rm{s}}\,\biggl(\frac{\eta_{\rm{j}}}{6.2\times10^{-4}}\biggl)^{-1}\biggl(\frac{\theta}{5^\circ}\biggl)^2\biggl(\frac{R_*}{5.8\times10^{13}\,\rm{cm}}\biggl).
\label{jet breakout time}
\end{eqnarray}

Substituting Eqs. (\ref{mass accretion rate}), (\ref{jet injection time}) and (\ref{jet breakout time}) into Eq. (\ref{cocoon energy1}), we get the formula for the cocoon energy,
\begin{eqnarray}
\Ec&=&2\eta_{\rm{j}}M_{\rm{core}}c^2\ln\biggl(\frac{t_{\rm{b}}}{t_{\rm{in}}}\biggl)
\label{cocoon energy2}\\
&=&2.0\times10^{56}\,{\rm{erg}}\biggl(\frac{\eta_{\rm{j}}}{6.2\times10^{-4}}\biggl)\biggl(\frac{M_{\rm{core}}}{9.6\times10^3\,\Msun}\biggl)\ln\biggl[\biggl(\frac{t_{\rm{b}}}{8.9\times10^3\,\rm{s}}\biggl)\biggl(\frac{t_{\rm{in}}}{9.7\times10^{-1}\,\rm{s}}\biggl)^{-1}\biggl].
\nonumber
\end{eqnarray}
This equation reproduces the cocoon energy shown in Table \ref{cocoon parameter table}, which are obtained by numerically integrating Eq. \eqref{cocoon energy1}.

Next, we also estimate the cocoon mass $\Mc$ at the jet breakout.
The cocoon mass $\Mc$ is defined as the mass in the cocoon when the jet head breaks out of the stellar surface.
Then, the cocoon mass is equal to the stellar mass within the volume where the cocoon expands in the progenitor.
We approximate the shape of the cocoon component as a cone whose height is the distance of the jet head from the stellar center, and whose radius is the distance of the cocoon surface from the jet axis.
We evaluate the cocoon mass as follows,
\begin{eqnarray}
\Mc&\simeq&\frac{M_*}{4\pi{R_*^3}/3}\times\frac{1}{3}\pi{R_c}^2R_*,
\label{cocoon mass1}
\end{eqnarray}
where $R_{\rm{c}}$ is the radius of the cocoon component at the jet breakout time.
In the second factor in Eq. (\ref{cocoon mass1}), we use the fact that when the jet head breaks out of the progenitor, the height of the cocoon is equal to the stellar radius $R_*$.

The radius is given by $R_{\rm{c}}\simeq{\beta_{\rm{c}}c}t_{\rm{b}}$, where $\beta_{\rm{c}}$ is the velocity of the cocoon component expanding in the progenitor star.
The velocity is estimated by the cocoon pressure $P_{\rm{c}}$ and the mean density in the cocoon component $\bar{\rho}$ as \citep{1989ApJ...345L..21B}
\begin{eqnarray}
\beta_{\rm{c}}\simeq\sqrt{\frac{P_{\rm{c}}}{\bar{\rho}c^2}}.
\label{cocoon velocity1}
\end{eqnarray}
Assuming that the cocoon is radiation-pressure-dominated, the cocoon pressure is given by $P_{\rm{c}}\simeq\Ec/\pi{r_{\rm{h}}}r_{\rm{c}}^2$.
The mean density of the cocoon component $\bar{\rho}$ is also approximated by $\bar{\rho}\simeq{3M_{r_{\rm{h}}}}/4\pi{r}_{\rm{h}}^3$.
Substituting the expression of the cocoon pressure and the mean density and Eqs. (\ref{mass coordinate}) and (\ref{cocoon energy2}) into Eq. (\ref{cocoon velocity1}), we obtain  
\begin{eqnarray}
\beta_{\rm{c}}\simeq\biggl(\frac{8\eta_{\rm{j}}}{3}\frac{\ln(t/t_{\rm{in}})}{1+3\ln(r_{\rm{h}}/R_{\rm{core}})}\biggl)^{1/2}\frac{\beta_{\rm{h}}ct}{r_{\rm{c}}}.
\label{cocoon velocity2}
\end{eqnarray}
Using the fact that the second factor in the parenthesis in Eq.(\ref{cocoon velocity2}) is of the order of $3.4\times10^{-1}$ at the jet breakout and that the cocoon radius is given by $r_{\rm{c}}\simeq{\beta_{\rm{c}}}ct$, we obtain the cocoon velocity as
\begin{eqnarray}
\beta_{\rm{c}}\simeq\biggl(\frac{8\eta_{\rm{j}}}{3}\frac{\ln(t_{\rm{b}}/t_{\rm{in}})}{1+3\ln(R_*/R_{\rm{core}})}\biggl)^{1/4}\beta_{\rm{h}}^{1/2}\simeq7.2\times10^{-2}\,\biggl(\frac{\eta_{\rm{j}}}{6.2\times10^{-4}}\biggl)^{3/4}\biggl(\frac{\theta}{5^\circ}\biggl)^{-1}.
\label{cocoon velocity3}
\end{eqnarray}
With Eq. (\ref{cocoon velocity3}), Eq. (\ref{cocoon mass1}) gives 
\begin{eqnarray}
M_{\rm{c}}&\simeq&\frac{1}{4}M_*\biggl(\frac{\beta_{\rm{c}}}{\beta_{\rm{h}}}\biggl)^2
\label{cocoon mass2}\\
&\simeq&2.8\times10^3\,\Msun\,\biggl(\frac{M_*}{10^5\,\Msun}\biggl)\biggl(\frac{\eta_{\rm{j}}}{6.2\times10^{-4}}\biggl)^{-1/2}\biggl(\frac{\theta}{5^\circ}\biggl)^2.
\nonumber
\end{eqnarray}
This roughly reproduces the numerical results in Table \ref{cocoon parameter table}.

\section{B. Light curve model}\label{Light curve model}
We describe one-zone analytical formulae for the light curves of cocoon emission based on \cite{1980ApJ...237..541A,1993ApJ...414..712P,2013ApJ...778...67N,2013ApJ...772...30D}.
We consider an expanding cocoon fireball which is non-relativistic and radiation-pressure-dominated.
Immediately after the breakout, the cocoon fireball is accelerated by the pressure $PdV$ work.
When its radius gets about doubled, we can assume that the cocoon energy is equally divided into the kinetic and the internal energy, $E_{\rm{kin}}\sim{E_{\rm{int}}}\sim{E_{\rm{c}}}/2$, and that the cocoon fireball starts homologous expansion $v \propto r$.
For simplicity, we also assume that the cocoon fireball is homogeneous.
In this phase, we can calculate light curves of the cocoon emission in the same way as Type \2P SNe, only with the progenitor's radius $R_*$, the cocoon energy $\Ec$, and mass $\Mc$ at the jet breakout.
Since the cocoon fireball expands with a constant velocity, the cocoon radius is given by
\begin{eqnarray}
\Rc(t)=\vc{t}+\Rc(0),
\label{cocoon radius}
\end{eqnarray}
where $\Rc(0)=2R_*$ and $\vc$ is the cocoon velocity which is evaluated by $\vc\sim\sqrt{5\Ec/3\Mc}$, where the numerical factor arises from the calculation of the total kinetic energy of the cocoon fireball ($E_{\rm{kin}}\simeq\int(\rho{}v^2/2){4\pi{r^2dr}}=3\Mc\vc^2/10$).
We should note that we set the origin of time at the moment when the cocoon fireball begins to expand homologously. 

We calculate the thermal evolution of the cocoon fireball with the first law of thermodynamics,
\begin{eqnarray}
\frac{d{E_{\rm{int}}}}{dt}=-P\frac{dV}{dt}+H-L,
\label{first law of thermodynamics}
\end{eqnarray}
where $E_{\rm{int}}$, $P$, $V=4\pi{R_{\rm{c}}}^3/3$, $H$, and $L$ are the total internal energy of the cocoon fireball, the cocoon pressure, the cocoon volume, the heating rate from the external energy source, and the radiative cooling rate, respectively.
In our study, we ignore the heating source of the cocoon fireball.
Since the cocoon fireball is radiation-pressure-dominated, the cocoon pressure $P$ is given by $P=E_{\rm{int}}/3V$.

First, we ignore the effect of hydrogen recombination on the opacity.
As long as the effective temperature of the cocoon fireball $T_{\rm{eff}}$ is higher than the recombination temperature $T_{\rm{ion}}\simeq6000\rm{\,K}$, we assume that the cocoon fireball is fully ionized.
Then, the cocoon fireball is optically thick and its luminosity is given by the diffusion approximation as,
\begin{eqnarray}
L&\simeq&4\pi{\Rc^2}\frac{cV}{3\kappa\Mc}\frac{E_{\rm{int}}}{V\Rc}=\frac{4\pi{c}\Rc}{3\kappa\Mc}E_{\rm{int}}\nonumber\\
&=&\frac{t+t_{\rm{e}}}{t_{\rm{d}}^2}E_{\rm{int}},
\label{diffusion equation}
\end{eqnarray}
where $\kappa$ is the cocoon opacity.
In Eq. (\ref{diffusion equation}), we define the expansion time $t_{\rm{e}}$ and the diffusion time $t_{\rm{d}}$ as follows,
\begin{eqnarray}
t_{\rm{e}}&:=&\frac{\Rc(0)}{\vc}\simeq1.1\times10^{4}\,{\rm{s}}\ {\Rc}_{,14}{\Ec}_{,56}^{-1/2}{\Mc}_{,3}^{1/2},
\label{expansion time}\\
\nonumber\\
t_{\rm{d}}&:=&\sqrt{\frac{3\kappa\Mc}{4\pi{\vc{c}}}}\simeq2.5\times10^{7}\,{\rm{s}}\ {\Ec}_{,56}^{-1/4}{\Mc}_{,3}^{3/4},
\label{diffusion time}
\end{eqnarray}
where ${\Rc}_{,14}=\Rc(0)/10^{14}\rm{\,cm}$, ${\Ec}_{,56}=\Ec/10^{56}\rm{\,erg}$, and ${\Mc}_{,3}=\Mc/10^{3}\Msun$.
In the second equality of Eq. (\ref{diffusion time}), we substitute the opacity value for $\kappa=0.35\,\rm{cm^{2}\,g^{-1}}$, which is the Thomson scattering opacity of the primordial chemical composition.
Using Eqs. (\ref{cocoon radius}), (\ref{first law of thermodynamics}), and (\ref{diffusion equation}), we obtain a differential equation for the cocoon luminosity,
\begin{eqnarray}
\frac{dL}{dt}+\frac{t+t_{\rm{e}}}{t_{\rm{d}}^2}L=0.
\label{luminosity equation}
\end{eqnarray}
Integration of Eq. (\ref{luminosity equation}) yields the time evolution of the luminosity,
\begin{eqnarray}
L(t)&=&\frac{t_{\rm{e}}\Ec}{2t_{\rm{d}}^2}\exp\biggl(-\frac{1}{2t_{\rm{d}}^2}(t^2+2t_{\rm{e}}t)\biggl)\simeq9.0\times10^{44}\,{\rm{erg\,s^{-1}}}\ {\Rc}_{,14}{\Ec}_{,56}{\Mc}_{,3}^{-1}\exp\biggl(-\frac{1}{2t_{\rm{d}}^2}(t^2+2t_{\rm{e}}t)\biggl).
\label{cocoon luminosity1}
\end{eqnarray}
When the exponential factor of this equation is almost unity during this phase ($t_{\rm{i}}^2/2t_{\rm{d}}^2\lesssim1$, see below), the bolometric luminosity dose not change so much.  
We equate Eq. (\ref{cocoon luminosity1}) with $L=4\pi\Rc(t)^2\sigma_{\rm{SB}}T_{\rm{eff}}(t)^4$, where $\sigma_{\rm{SB}}$ is the Stefan-Boltzmann constant, and get the time evolution of the effective temperature as 
\begin{eqnarray}
T_{\rm{eff}}(t)\simeq\biggl(\frac{c\Rc(0)}{10\kappa\sigma_{\rm{SB}}(t+t_{\rm{e}})^2}\biggl)^{1/4}\exp\biggl(-\frac{1}{8t_{\rm{d}}^2}(t^2+2t_{\rm{e}}t)\biggl).
\label{effective temperature}
\end{eqnarray}
In this phase, the photospheric velocity $v_{\rm{ph}}$ is roughly evaluated by the cocoon velocity $\vc$.
Then using the definition of the cocoon velocity, we obtain
\begin{eqnarray}
v_{\rm{ph}}\sim\biggl(\frac{5\Ec}{3\Mc}\biggl)^{1/2}\simeq3.0\times10^{-1}c\ {\Ec}_{,56}^{1/2}{\Mc}_{,3}^{-1/2}.
\label{photospheric velocity}
\end{eqnarray}

Next, we take the recombination effect into account.
When the effective temperature gets smaller than the critical value, a recombination wave starts to recede into the center.
We define the time $t_{\rm{i}}$ as the moment when the effective temperature drops to the critical one $T_{\rm{eff}}(t_{\rm{i}})=T_{\rm{ion}}$. 
Then, the time $t_{\rm{i}}$ is given by
\begin{eqnarray}
t_{\rm{i}}\simeq\biggl(\frac{c\Rc(0)}{10\kappa\sigma_{\rm{SB}}}\biggl)^{1/2}T_{\rm{ion}}^{-2}\simeq3.4\times10^{6}{\rm{\,s}}\ {\Rc}_{,14}^{1/2},
\label{recombination time}
\end{eqnarray}
where we assume that $t_{\rm i} \gg t_{\rm e}$ and the exponential factor in Eq. (\ref{effective temperature}) is unity.
These assumptions are justified for the cocoon parameters we consider ($t_{\rm e}/t_{\rm i} \ll 1$ and $t_{\rm{i}}^2/8t_{\rm{d}}^2\ll1$).
Since it is transparent outside of the recombination wave, we identify the photospheric radius $R_{\rm{ph}}(t):=x_{\rm{i}}(t)\Rc(t)$ as the recombination front, where the temperature is equal to the recombination temperature.
We also consider the thermal evolution of the cocoon fireball with Eq. (\ref{first law of thermodynamics}) only for the volume within the photospheric radius.
Then, the internal energy within the photospheric radius is given by $\tilde{E}_{\rm{int}}=E_{\rm{int}}\tilde{V}/V$, where $\tilde{V}:=x_{\rm{i}}(t)^3V$ is the volume within the photospheric radius.
The luminosity is given by $L=4\pi{R_{\rm{ph}}(t)}^2\sigma_{\rm{SB}}T_{\rm{ion}}^4$.
The diffusion approximation also gives the luminosity as
\begin{eqnarray}
L\simeq4\pi{x_{\rm{i}}(t)^2\Rc(t)^2}\frac{Vc}{3\kappa\Mc}\frac{\tilde{E}_{\rm{int}}}{\tilde{V}x_{\rm{i}}(t)\Rc(t)}=\frac{4\pi{c}}{3\kappa\Mc}\frac{\Rc(t)}{x_{\rm{i}}(t)^2}\tilde{E}_{\rm{int}}.
\label{cocoon luminosity2}
\end{eqnarray} 
Equating these expressions, we obtain the internal energy within the photospheric radius as
\begin{eqnarray}
\tilde{E}_{\rm{int}}=\frac{3\kappa\Mc\sigma_{\rm{SB}}T_{\rm{ion}}^4}{c}\Rc(t)x_{\rm{i}}(t)^{4}.
\label{cocoon internal energy}
\end{eqnarray}
With Eqs. (\ref{cocoon luminosity2}) and (\ref{cocoon internal energy}), the first law of thermodynamics yields a differential equation for $x_{\rm{i}}(t)$ as 
\begin{eqnarray}
\frac{dx_{\rm{i}}}{dt}=-\frac{2x_{\rm{i}}}{5t}-\frac{t}{5t_{\rm{d}}^2x_{\rm{i}}},
\label{equation for xi}
\end{eqnarray}
where we approximate the cocoon radius as $\Rc(t)\sim{\vc}t$ ($t>t_{\rm{i}}\gg{t}_{\rm{e}}$).
We integrate Eq. (\ref{equation for xi}) with the initial condition $x_{\rm{i}}(t_{\rm{i}})=1$, and obtain the time evolution of the photospheric radius
\begin{eqnarray}
R_{\rm{ph}}(t)^2=x_{\rm{i}}(t)^2\Rc(t)^2=\vc^2\biggl[t^{6/5}t_{\rm{i}}^{4/5}\biggl(1+\frac{t_{\rm{i}}^2}{7t_{\rm{d}}^2}\biggl)-\frac{t^4}{7t_{\rm{d}}^2}\biggl]\,\,\,\,\,(t>{t_{\rm{i}}}).
\label{photospheric radius in recombination}
\end{eqnarray}

We summarize the time evolutions of the bolometric luminosity, the effective temperature, and the photospheric radius as,
\begin{eqnarray}
L(t)&=&\begin{cases}
\frac{t_{\rm{e}}\Ec}{2t_{\rm{d}}^2}\exp\biggl(-\frac{t^2}{2t_{\rm{d}}^2}\biggl)&\text{$(t_{\rm{e}}\ll{t}<t_{\rm{i}})$}\\
4\pi\sigma_{\rm{SB}}T_{\rm{ion}}^4\vc^2\biggl[t^{6/5}t_{\rm{i}}^{4/5}\biggl(1+\frac{t_{\rm{i}}^2}{7t_{\rm{d}}^2}\biggl)-\frac{t^4}{7t_{\rm{d}}^2}\biggl]&\text{$(t>t_{\rm{i}}),$}
\end{cases}
\label{summary bolometric luminosity}\\
T_{\rm{eff}}(t)&=&\begin{cases}
\biggl(\frac{cR(0)}{10\kappa\sigma_{\rm{SB}}}\biggl)^{1/4}t^{-1/2}\exp(-\frac{t^2}{8t_{\rm{d}}^2})&\text{$(t_{\rm{e}}\ll{t}<t_{\rm{i}})$}\\
T_{\rm{ion}}&\text{$(t>t_{\rm{i}}),$}
\end{cases}
\label{summary effective temperature}\\
\Rph(t)&=&\begin{cases}
\vc{t}&\text{$(t_{\rm{e}}\ll{t}<t_{\rm{i}})$}\\
\vc\biggl[t^{6/5}t_{\rm{i}}^{4/5}\biggl(1+\frac{t_{\rm{i}}^2}{7t_{\rm{d}}^2}\biggl)-\frac{t^4}{7t_{\rm{d}}^2}\biggl]^{1/2}
&\text{$(t>t_{\rm{i}}).$}
\end{cases}
\label{summary photospheric radius}
\end{eqnarray}
The duration of the cocoon emission is given by solving $L(\Delta{t_{\rm{co}}})=0$.
We obtain the duration as
\begin{eqnarray}
\Delta{t_{\rm{co}}}=7^{5/14}t_{\rm{i}}^{2/7}t_{\rm{d}}^{5/7}\simeq2.8\times10^{7}{\rm{\,s}}\ {\Rc}_{,14}^{1/7}{\Ec}_{,56}^{-5/28}{\Mc}_{,3}^{15/28}.
\label{plateau duration}
\end{eqnarray}
The time $t_{\rm{2nd}}$ when the light curves show the second peak is also given by solving $dL/dt=0$.
We get the time as 
\begin{eqnarray}
t_{\rm{2nd}}=\biggl(\frac{21}{10}\biggl)^{5/14}t_{\rm{i}}^{2/7}t_{\rm{d}}^{5/7}\simeq1.9\times10^{7}{\rm{\,s}}\ {\Rc}_{,14}^{1/7}{\Ec}_{,56}^{-5/28}{\Mc}_{,3}^{15/28}.
\label{peak time2}
\end{eqnarray}
Then, the bolometric luminosity reaches 
\begin{eqnarray}
L_{\rm{2nd}}=4\pi\sigma_{\rm{SB}}T_{\rm{ion}}^4\vc^2\biggl[t_{\rm{max}}^{6/5}t_{\rm{i}}^{4/5}\biggl(1+\frac{t_{\rm{i}}^2}{7t_{\rm{d}}^2}\biggl)-\frac{t_{\rm{max}}^4}{7t_{\rm{d}}^2}\biggl]\simeq4.7\times10^{45}{\rm{\,erg\,s^{-1}}}\ {\Rc}_{,14}^{4/7}{\Ec}_{,56}^{11/14}{\Mc}_{,3}^{-5/14}.
\label{peak luminosity2}
\end{eqnarray}
The photospheric velocity at this moment is expressed by 
\begin{eqnarray}
v_{\rm{ph}}(t_{\rm{2nd}})=x_{\rm{i}}(t_{\rm{2nd}})\vc\simeq3.5\times10^{9}{\rm{\,cm\,s^{-1}}}\ {\Rc}_{,14}^{1/7}{\Ec}_{,56}^{4/7}{\Mc}_{,3}^{-5/7}.
\label{photospheric velocity2}
\end{eqnarray}

The observed flux of the cocoon emission is given by 
\begin{eqnarray}
F_{\lambda_{\rm{obs}}}(t_{\rm{obs}})d\lambda_{\rm{obs}}=\pi{B_{\lambda}(T_{\rm{eff}}(t))}\frac{R_{\rm{ph}}(t)^2}{d_{\rm{L}}^2}d\lambda,
\label{flux}
\end{eqnarray}
where $B_{\lambda}(T)=2hc^2/\lambda^5(\exp(hc/\lambda{k_{\rm{B}}}T)-1)^{-1}$ and $d_{\rm{L}}$ are the Planck function and the luminosity distance, respectively.
The time in the observer frame $t_{\rm{obs}}$ and the observed wavelength $\lambda_{\rm{obs}}$ are related with the time and the wavelength in the progenitor rest frame as $t_{\rm{obs}}=(1+z)t$ and $\lambda_{\rm{obs}}=(1+z)\lambda$.

{


\end{document}